\documentclass[preprint]{elsarticle}

\biboptions{sort&compress}
\usepackage{amssymb}
\usepackage{graphicx}
\usepackage{pdfpages}
\usepackage{adjustbox}
\usepackage{rotating}
\usepackage[T1]{fontenc}
\usepackage{mathrsfs}
\usepackage{multirow}
\usepackage{hhline}
\usepackage{enumerate}
\usepackage{longtable}
\usepackage{enumitem}
\usepackage{booktabs}
\usepackage[lined,commentsnumbered]{algorithm2e}
\usepackage{algpseudocode}
\usepackage{hvfloat}
\usepackage[english]{babel}
\usepackage{stackengine}
\let\emptyset\varnothing

\newlength\marincrease


\usepackage[font=footnotesize,caption=false]{subfig}
\usepackage{lscape}
\usepackage{mathtools}
\usepackage{etoolbox}
\usepackage{framed}
\usepackage{xcolor}[2007/01/21]
\usepackage{listings}

\definecolor{mygreen}{rgb}{0,0.6,0}
\definecolor{mygray}{rgb}{0.5,0.5,0.5}
\definecolor{mymauve}{rgb}{0.58,0,0.82}
\lstdefinestyle{customc}{
  belowcaptionskip=1\baselineskip,
  breakatwhitespace=false,         
  breaklines=true,                 
  captionpos=b,                    
  commentstyle=\color{mygreen},    
  deletekeywords={...},            
  escapeinside={\%*}{*)},          
  extendedchars=true,      
  frame=single,	                   
  language=C,
  showstringspaces=false,
  basicstyle=\footnotesize\ttfamily,
  keywordstyle=\bfseries\color{green!40!black},
  commentstyle=\itshape\color{purple!40!black},
  identifierstyle=\color{blue},
  stringstyle=\color{orange},
}

 \usepackage{hyperref}
\usepackage[
 open,
 openlevel=2,
 atend
 ]{bookmark}[2011/12/02]

 \bookmarksetup{color=blue}
 \makeatletter
  \let\old@@children\@@children
  \def\@@children{\futurelet\my@next\my@@children}
  \def\my@@children{%
  \ifx\my@next\missing\else
  \expandafter\@gobble
  \fi
  \expandafter\old@@children}
  
 \makeatother
 
 \newcommand{\missing}{ \edge[draw=none]; {} }
 
 \providecommand{\floor}[1]{\left \lfloor #1 \right \rfloor }

 \usepackage{makecell}
 
\begin{document}

\renewcommand{\thesection}{\Roman{section}}

\newtheorem{theoremm}{Theorem}
\newtheorem{eqed}{Example}
\newtheorem {lemmaa}{Lemma}
\newtheorem{proposition}{Proposition}
\newtheorem {observation}[theoremm]{Observation}
\newtheorem {defnn}{Definition}
\newtheorem {corollaryy}{Corollary}
\newtheorem {conjecturee}{Conjecture}
\newtheorem {fact}[theoremm]{Fact}
\newtheorem {procd}{Procedure}
\newtheorem {rules}{Rule}
\newenvironment{example}{\begin{eqed} \rm}{\hfill\end{eqed}}
\newenvironment{proof}{\noindent {\bf Proof :\ } }{\hfill $\Box$ }
\newenvironment{lemma}{\begin{lemmaa} \sl}{\end{lemmaa}}
\newenvironment{theorem}{\begin{theoremm}{\bf :}\sl}{\end{theoremm}}
\newenvironment{corollary}{\begin{corollaryy}{\bf :}\sl}{\end{corollaryy}}
\newenvironment{procd1}{\begin{procd} \sl}{\end{procd}}
\newenvironment{conjecture}{\begin{conjecturee} \sl}{\end{conjecturee}}
\newenvironment{definition}[1][Definition]{\begin{defnn} \sl}{\end{defnn}}

\begin{frontmatter}

\title{On Finite $1$-Dimensional Cellular Automata: Reversibility and Semi-reversibility}


\author[mymainaddress]{Kamalika Bhattacharjee\corref{mycorrespondingauthor}}
\cortext[mycorrespondingauthor]{Corresponding author}
\ead{kamalika.it@gmail.com}
%
%
\author[mymainaddress]{Sukanta Das}
\ead{sukanta@it.iiests.ac.in}
\address[mymainaddress]{Department of Information Technology, Indian Institute of Engineering Science and Technology, Shibpur, West Bengal, India 711103}

\begin{abstract}
Reversibility of a one-dimensional finite cellular automaton (CA) is dependent on lattice size.
A finite CA can be reversible for a set of lattice sizes. On the other hand, reversibility of an infinite CA, which is decided by exploring the rule only, is different in its kind from that of finite CA. 
Can we, however, link the reversibility of finite CA to that of infinite CA?
In order to address this issue, we introduce a new notion, named \emph{semi-reversibility}. We classify the CAs into three types with respect to reversibility property -- reversible, semi-reversible and strictly irreversible. A tool, {\em reachability tree}, has been used to decide the reversibility class of any CA. Finally, relation among the existing cases of reversibility is established.

\end{abstract}

\begin{keyword}
Cellular Automata (CAs)\sep Rule Min Term (RMT)\sep Reversibility \sep Semi-reversibility\sep Strictly Irreversibility\sep Reachability Tree
\end{keyword}

\end{frontmatter}


\section{Introduction}
\label{Chap:semireversible:sec:intro}
\noindent The reversibility property of a cellular automaton (CA) refers to that each configuration of the CA has a unique predecessor. This implies, there is no loss of information during the evolution of the CA. It has direct correspondence with the reversibility of microscopic physical systems, implied by the laws of quantum mechanics. The reversible (or bijective) cellular automata (CAs) have been utilized in different domains, like simulation of natural phenomenon \cite{hartman90}, cryptography \cite{ppc1}, pattern generations \cite{Kari2012180}, pseudo-random number generation \cite{aspdac04}, recognition of languages \cite{Kutrib20081142} etc. In this work, we are concentrating on reversibility for $1$-dimensional CAs.

Classically cellular automata (CAs) are defined over infinite lattice. In classical literature, therefore, reversibility of a CA simply implies the reversibility of the CA, defined over infinite lattice. On the other hand, the issue of reversibility  of finite CAs has gained the interest of researchers in last few decades due to various reasons. For finite CAs, lattice size along with a CA rule is a parameter to decide its reversibility. Hence, there is an ambiguity in the term ``reversibility of a CA'', because the CA can be defined over finite as well as infinite lattice.

Additionally, reversibility of CAs under finite and infinite cases are two different issues which are not to be mixed up. There exist classical algorithms (e.g. \cite{Amoroso72,suttner91}) to decide reversibility of $1$-dimensional infinite CAs. However, these classical algorithms do not work for finite CAs. Therefore, there is no apparent bridge between finite and infinite cases. The primary motivation of this paper is to inquire of a relation between these two cases.


\subsection{Some Existing Results on Reversibility}
\noindent While studying the reversibility (i.e. injectivity) of infinite and finite CAs over a given local map $R$, there are at least the following four cases:
\begin{description} 
\item[Case $1$:]\label{case1} An infinite CA whose global function is injective on the set of ``all infinite configurations''.
\item[Case $2$:]\label{case2} An infinite CA whose global function is injective on the set of ``all {\em periodic} configurations''. In one-dimension, a configuration $x$ is periodic, or more precisely, spatially periodic if there exists $p \in \mathbb{N}$ such that $x_{i+p}=x_i$ for all $i\in \mathbb{Z}$.
\item[Case $3$:]\label{case3} An infinite CA whose global function is injective on the set of ``all finite configurations of length $n$'' for all $n\in \mathbb{N}$.
\item[Case $4$:]\label{case4} A finite CA whose global function is injective on the set of ``all configurations of length $n$'' for a fixed $n$.
\end{description} 

Therefore, for a CA with a given local map $R$, there exists at least four types of global transition functions depending on the above four cases -- let, $G$ be the global transition function on the set of all infinite configurations, $G_P$ be the global transition function on the set of periodic configurations, $G_F$ be the global transition function on the set of finite configurations for all lattice size $n \in \mathbb{N}$ and $G_n$ be the global transition function over a fixed lattice size $n$. For one-dimensional CAs, periodic boundary condition over configurations of length $n$, for all $n \in \mathbb{N}$ evidently implies periodic configurations. So, in this paper, by the global transition function on the set of periodic configurations, we mean both the notions and use result on any one of these to portray the other. 

There are many interesting results regarding reversibility of $1$-dimensional CAs, over a given local map $R$, for the first three cases: 
\begin{theorem}
The following statements hold for $1$-dimensional infinite CAs:
\begin{enumerate}[topsep=0pt,itemsep=0ex,partopsep=2ex,parsep=1ex]
\item A CA, defined over infinite lattice, is reversible, if inverse of its global transition function $G$, $G^{-1}$ is the global transition function of some (infinite) CA \cite{hedlund69}.

\item An infinite CA is reversible, if $G$ is injective \cite{hedlund69,Richa72}.

\item If $G$ is injective, then $G_F$ is surjective \cite{Richa72}.

\item $G_F$ is injective if and only if $G$ is surjective \cite{moore1962machine,Myhill63}.

\item $G_F$ is surjective if and only if it is bijective \cite{amoroso1970garden}.

\item If $G_F$ is surjective, then $G_F$ is injective; but the converse is not true \cite{Richa72}.

\item  $G$ is injective, if and only if $G_P$ is injective \cite{sato77}.


\item If $G_P$ is injective, then $G$ is surjective \cite{sato77}.

\item $G_P$ is surjective, if and only if $G$ is surjective \cite{sato77}.

\item If $G_P$ is injective, then $G_P$ is surjective \cite{sato77}.

\item If $G_P$ or $G_F$ is surjective, then $G$ is surjective \cite{Kari05}.

\item If $G$ is injective, then $G_P$ and $G_F$ are injective \cite{Kari05}.

\item If $G_P$ is injective, then $G$ is injective \cite{Kari05}.

\item If $G$ is surjective, then $G_P$ is surjective \cite{Kari05}.
\end{enumerate}
\end{theorem}

\begin{figure}[!h]
\centering
\includegraphics[width= 2.5in, height = 0.8in]{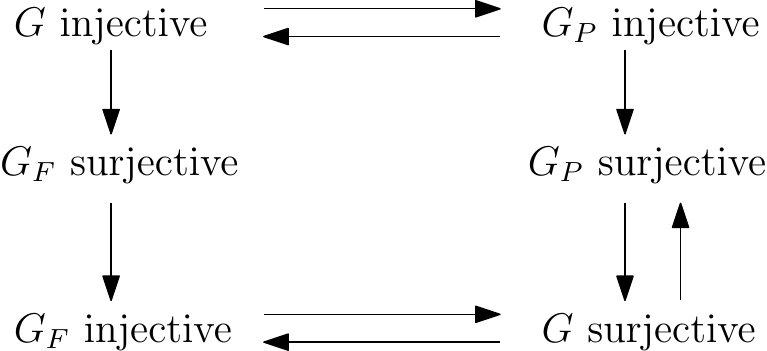}
\caption{Relationship between Injectivity and Surjectivity properties of $1$-dimensional CAs}
\label{Chap:semireversible:fig:rev_rel}
\end{figure}

\noindent The relation among $G$, $G_F$ and $G_P$ is shown in Figure~\ref{Chap:semireversible:fig:rev_rel}. Therefore, for a one dimensional CA with local rule $R$, injectivity of $G$ implies injectivity of $G_F$ and $G_P$. However, Case $4$, that is, reversibility for finite CAs, where $G_n$ is injective over the set of all configurations of length $n$ for a fixed $n$, is different from these three cases. In \cite{jca2015}, an algorithm is presented which can decide reversibility of a $d$-state $3$-neighborhood finite CA for a given lattice size $n$.  

\subsection{The Challenges}
\noindent One can observe that, there is a both way implication between the injectivity of $G_P$ and $G$ (see Figure~\ref{Chap:semireversible:fig:rev_rel}). That is, if a CA is injective over the set of all periodic configurations, it is also injective over the set of all infinite configurations and vice versa. Further, injectivity of an infinite CA implies reversibility of it. Hence, one can exploit the injectivity (or, bijectivity) of $G_P$ over periodic configurations to deduce about injectivity (or, bijectivity) of $G$.

Moreover, injectivity of $G_P$ implies injectivity of the CA over configurations of length $n$ under periodic boundary condition for all $n \in \mathbb{N}$. But, injectivity of CAs over configurations of length $n$ under periodic boundary condition is the Case $4$, where the global transition function is $G_n$. Hence, by using injectivity of $G_n$ for all $n \in \mathbb{N}$, we may extrapolate about injectivity of $G_P$. That is, by exploiting reversibility of finite CAs with discrete $n$, we are targeting to infer about reversibility of CAs defined over infinite lattice. 

But problem with this approach is, injectivity of $G_n$ (Case $4$) is dependent on the lattice size $n$. That is, for a Case $4$ CA where number of states per cell is $d$ and size of the neighborhood is $m$, we need to additionally know the lattice size to decide its reversibility.
In fact, for many of these CAs, the set of discrete lattice sizes for which the CA is reversible is an infinite set. For instance, elementary CA (ECA) $101$ is reversible if $n$ is odd, but ECA $51$ is reversible for any $n \in \mathbb{N}$. In \cite{Ino05,Sato2009, 0305-4470-37-22-006}, existence of elementary CAs (ECAs) and $2$-neighborhood $3$-state CAs which are reversible for infinite number of lattice sizes under periodic boundary condition are explored. 
However, classical reversibility of the infinite CAs referred by Case $1$ as well as of Case $2$ and $3$ is independent of lattice size. So, to conclude about injectivity of $G_P$, and correspondingly about $G$ and $G_F$, $G_n$ has to be injective for every $n \in \mathbb{N}$.

There are many finite CAs (Case $4$) which are reversible for some $n$; that is, for a given local rule $R$, $G_n$ is bijective for those $n$.
However, as $G_n$ is not bijective for all $n \in \mathbb{N}$, $G_P$ is not bijective for those CAs. So, in classical sense, these CAs are irreversible when defined over infinite lattice, but reversible when its lattice size is finite. Therefore, when CA is defined over finite lattice, we get a new class of reversibility -- reversible for some $n \in \mathbb{N}$. To deal with this class of CAs, we introduce the notion of \emph{semi-reversibility}. Hence, there are three types of finite CAs -- (1) CAs for which $G_n$ is injective for each $n \in \mathbb{N}$, (2) CAs for which $G_n$ is injective for some $n \in \mathbb{N}$, and (3) CAs for which $G_n$ is not injective for each $n \in \mathbb{N}$.

The algorithm of \cite{jca2015} can decide reversibility of a finite CA with a local rule $R$ for a given lattice size $n$. However, for a finite CA with a set of (possibly infinite) lattice sizes, one needs to check every lattice size $n$, to see whether the CA is reversible/irreversible for each $n$ or some of these $n\in \mathbb{N}$. Nonetheless, if the set is not very small, it is very difficult to do such test.
In this scenario, the following two questions arise -- 
\begin{enumerate}
\item Is it possible to understand the reversibility behavior of a finite CA by exploring the CA for some small, but sufficient lattice size $n$? 

\item What is the relation between reversibility of finite CA over fixed $n$ (Case $4$) to that of the other three cases?
\end{enumerate}

This paper targets to answer these two questions for $1$-dimensional $d$-state $m$-neighborhood CAs under periodic boundary condition. Our contributions are mainly the following:
\begin{enumerate}
\item Identification of \emph{semi-reversible} CAs, which are classically treated as irreversible, but are reversible for a set (possibly infinite) of lattice sizes. Hence, we get a new classification of CAs -- reversible, semi-reversible and \emph{strictly} irreversible (Section~\ref{Chap:semireversible:sec:rev_class}).
\item A tool named \emph{reachability tree} is used to find the class of reversibility of a $1$-dimensional CA. Hence, we decide the reversibility behavior of a CA for any $n \in \mathbb{N}$ from a finite size (Section~\ref{Chap:semireversible:sec:rtree}).
\item Development of a scheme to decide (semi-)reversibility of a CA along with an expression to find the infinite set of sizes for which it is reversible (Section~\ref{Chap:semireversibility:sec:semi_tree}).
\item Establishment of relation among the four cases of reversibility (Section~\ref{Chap:semireversible:sec:remark}).
\end{enumerate}

\section{Definitions: Basics}\label{sec:CAbasic}
\noindent In this work, we consider $1$-dimensional CA where the cells are arranged as a ring $\mathscr{L}=\mathbb{Z}/n\mathbb{Z}$ and each cell can assume any of the states of $S = \{0,1, \cdots, d-1\}$. The cells change their states depending on the present states of itself, its $l_r$ number of consecutive left neighbors and $r_r$ number of consecutive right neighbors.
Therefore, the next-state function or local rule of the CA is $R: S^m \rightarrow S$, where $m= l_r + r_r + 1$. The local rule, or simply \textit{rule}, however, can be expressed by a tabular form (see Table~\ref{rule}). 
Such a table has an entry for each $m$-tuple -- $\langle s_0,s_1,\cdots,s_{m-1}\rangle$, corresponding to the $m$ neighbors of a cell.
This tuple is called as \textit{Rule Min Term (RMT)} \cite{jca2015}. 

 \begin{definition}
 Let $R: S^m \rightarrow S$ be a rule, where $m= l_r + r_r + 1$. An input $\langle s_0,s_1,\cdots,s_{m-1}\rangle$ to $R$ is called as {\em Rule Min Term (RMT)}.
 \end{definition}
 
 \begin{table}[h]
  \setlength{\tabcolsep}{1.3pt}
  \begin{center}
  \vspace{-1.2em}	
  \caption{Some rules for $d=3$ and $m=3$.}
  \label{rule}
  \resizebox{1.00\textwidth}{!}{
  \begin{tabular}{cccccccccccccccccccccccccccc}
  \toprule
 {\thead{RMT}} & \thead{222} & \thead{221} & \thead{220} & \thead{212} & \thead{211} & \thead{210} & \thead{202} & \thead{201} & \thead{200} & \thead{122} & \thead{121} & \thead{120} & \thead{112} & \thead{111} & \thead{110} & \thead{102} & \thead{101} & \thead{100} & \thead{022} & \thead{021} & \thead{020} & \thead{012} & \thead{011} & \thead{010} & \thead{002} & \thead{001} & \thead{000}\\ 
  
  & \thead{(26)} & \thead{(25)} & \thead{(24)} & \thead{(23)} & \thead{(22)} & \thead{(21)} & \thead{(20)} & \thead{(19)} & \thead{(18)} & \thead{(17)} & \thead{(16)} & \thead{(15)} & \thead{(14)} & \thead{(13)} & \thead{(12)} & \thead{(11)} & \thead{(10)} & \thead{(9)} & \thead{(8)} & \thead{(7)} & \thead{(6)} & \thead{(5)} & \thead{(4)} & \thead{(3)} & \thead{(2)} & \thead{(1)} & \thead{(0)}\\ 
   \midrule
 
   & 0&1&2&0&1&2&1&2&0&0&1&2&2&1&0&1&0&2&2&0&1&0&2&1&1&0&2\\
  & 2&0&1&0&1&2&2&1&0&2&0&1&0&1&2&2&1&0&2&0&1&0&1&2&2&1&0\\
  $\mathbf{R}$ & 2&2&2&2&1&1&1&1&2&0&0&1&0&0&0&0&0&0&1&1&0&1&2&2&2&2&1\\
   & 1&0&2&2&2&1&0&1&0&1&0&2&2&2&1&0&1&0&1&0&2&2&2&1&0&1&0\\
   &1&1&2&2&2&1&0&1&0&1&1&2&2&2&1&0&0&0&1&1&2&2&2&1&0&0&0\\
  \bottomrule
  \end{tabular}
  }
  \end{center}
  \vspace{-1.0em}
  \end{table}

Each RMT is associated to a number $r = s_0 \times d^{m-1} + s_1 \times d^{m-2} + \cdots + s_{m-2} \times d + s_{m-1}$, whereas $R(s_0,s_1,\cdots,s_{m-1})$ is presented by $R[r]$. 
The number of possible RMTs for a CA is $ d^{m} $.  In this work, we present a rule in two ways -- by a string ``$R[d^m-1] \cdots R[1]R[0]$'', or by its decimal equivalent. In case of ECAs, decimal numbers are used to present rules. 
%
\begin{definition}
   \label{dbg_def}
   Let $\Sigma$ be a set of symbols, and $k \geq 1$ be a number. Then, the de Bruijn graph is
   $B(k, \Sigma ) = (V, E)$, where $V = \Sigma^k$ is the set of vertices, and $E = \{(ax, xb)|a, b \in \Sigma, x \in \Sigma^{k-1}\}$ is the set of edges.
\end{definition}

\begin{figure}[!h]
 \centering
  \includegraphics[width=3.5in, height = 2.0in]{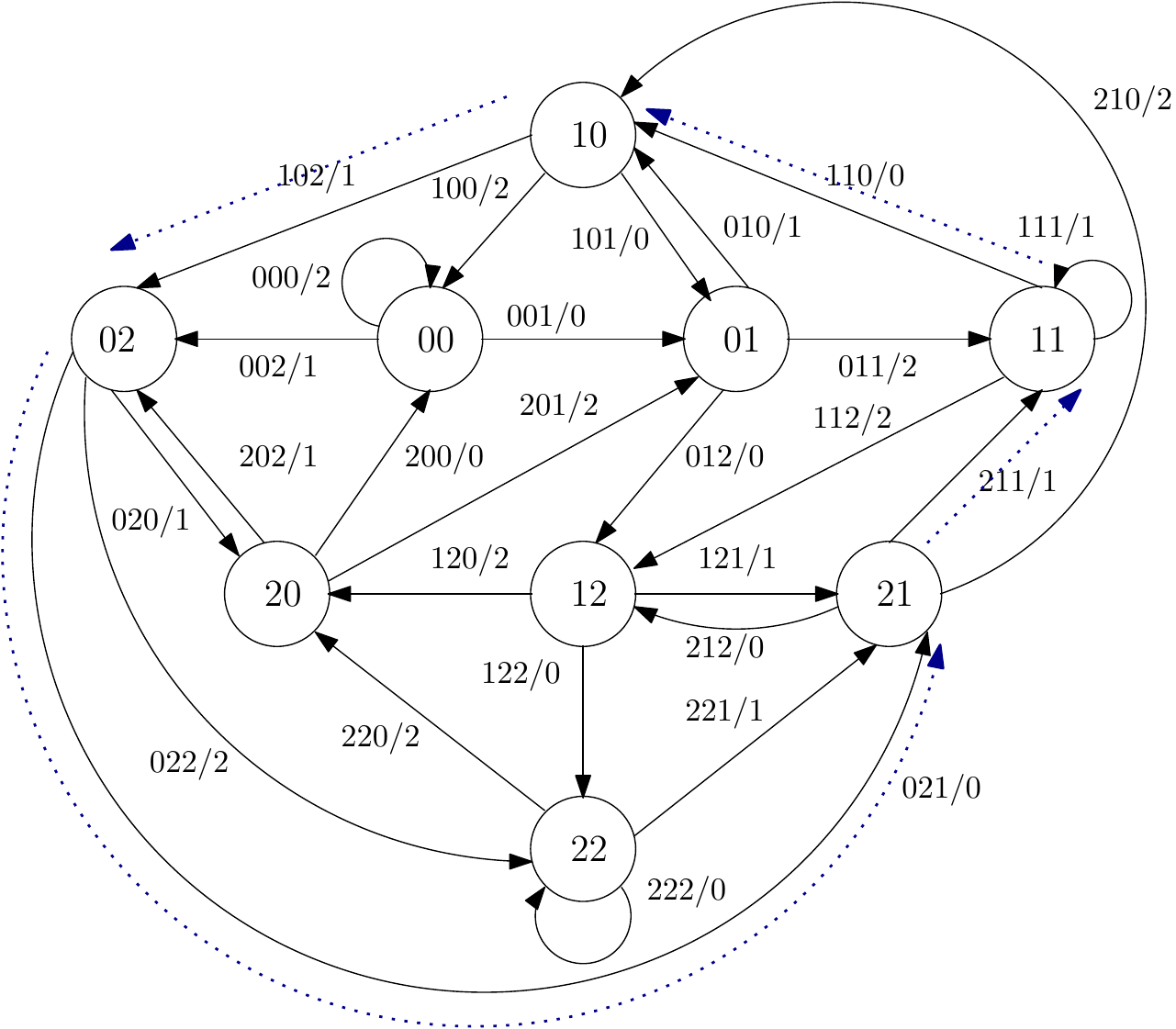}
    \caption{de Bruijn graph for a $3$-neighborhood $3$-state CA $012012120012210102201021102$ }                    \label{fig:dbg}     
\vspace{-1.9em} 
 \end{figure}

A one-dimensional CA can also be represented by a de Bruijn graph $B(m-1,S)$, where each edge $(ax, xb)$ of $B(m-1, S)$ depicts the overlapping sequence of nodes and $(axb) \in S^m$ \cite{suttner91}. Observe that, this overlapping sequence is equivalent to RMT. This graph has $d^{m-1}$ number of nodes and $d^m$ number of edges and is balanced in the sense that each vertex has both in-degree and out-degree $k$. 
Figure~\ref{fig:dbg} represents a $3$-neighborhood $3$-state CA $012012120012210102201021102$. For ease of understanding, each of the edges $(ax, xb)$ is labeled by the overlapping sequence of the connecting nodes $(axb)$ / $R(axb)\in S$.
   
\begin{definition}
\label{Def:balancedrule}
	A rule $R: S^m \rightarrow S$ is called \textbf{balanced} if it contains $d^{m-1}$ number of RMTs for each of the $d$ states possible for that CA; otherwise it is an \textbf{unbalanced} rule. 
	
\end{definition}

For example, the $3$-neighborhood $3$-state CA $012012120012210102201021102$ ($3^{rd}$ row of Table~\ref{rule}) is balanced, but the CA $112221010112221000112221000$ ($7^{th}$ row of Table~\ref{rule}) is unbalanced. 

A configuration of a CA is a mapping $c:\mathscr{L}\rightarrow S$, which specifies the states of all cells. That means, if $x$ is a configuration of an $n$-cell CA, then $x=(x_i)_{i\in\mathscr{L}}$, where $x_i$ is the state of cell $i$. During evolution, a CA hops from one configuration to another.
Let $\mathcal{C}_n$ be the set of all configurations of an $n$-cell CA and $G_n:\mathcal{C}_n\rightarrow \mathcal{C}_n$ be its global transition function induced by the local rule $R$. Then, 
\begin{equation*}
y=G_n(x) = G_n(x_0x_1\cdots x_{n-1})= (R(x_{i-l_r},\cdots, x_i, \cdots, x_{i+r_r}))_{i\in \mathscr{L}}
\end{equation*}
Here, $y$ is the successor of the configuration $x$. However, an alternative way of representing a configuration is \emph{RMT sequence}.

\begin{definition}
Let $x=(x_i)_{i\in\mathscr{L}}$ be a configuration of a CA. The \textbf{\em RMT sequence} of $x$, denoted as $\tilde{x}$, is $(r_i)_{i\in\mathscr{L}}$ where $r_i$ is the RMT $(x_{i-l_r},\cdots x_i, \cdots, x_{i+r_r})$.
\end{definition}

We can also get an RMT sequence from a de Bruijn graph by traversing a {cycle} in the graph. 
For instance, let $1021$ be a configuration of a $3$-neighborhood $3$-state CA with $4$ cells. Then the RMT sequence corresponding to this configuration is a cycle of length $4$ in the de Bruijn graph -- $\langle12(110), 11(102), 7(021), 22(211)\rangle$ (shown in directed dotted line in Figure~\ref{fig:dbg}). The next configuration for any configuration can, therefore, be found by traversing the de Bruijn graph. For example, the next configuration of the configuration $1021$ for the CA of Figure~\ref{fig:dbg} is $0101$.

However, only a specific set of RMTs can be chosen after an RMT in an RMT sequence. For example, in an RMT sequence of $3$-neighborhood $3$-state CAs, after RMT $12 (110)$, either of the RMTs $9 (100)$, $10 (101)$ or $11(102)$ can exist. The set of RMTs, from which the next RMT in an RMT sequence is to be chosen, is called \emph{sibling} RMTs. Similarly, the set of RMTs for which the next RMT is from the same sibling RMTs, is termed as \emph{equivalent} RMTs. In the de Bruijn graph of a CA, $d$ number of outgoing RMTs corresponding to each node are siblings to each other and $d$ number of incoming RMTs to that node are equivalents to each other \cite{jca2015}.

\begin{definition}
\label{def3}
A set of $d$ RMTs $r_1, r_2, ..., r_d$ of a $d$-state CA rule are said to be equivalent to each other if $r_1 d \equiv r_2 d \equiv ... \equiv r_d d \pmod{ d^m}$.

\end{definition}

\begin{definition}
\label{def4}
A set of $d$ RMTs $r'_1, r'_2, ..., r'_d$ of a $d$-state CA rule are said to be sibling to each other if $\floor{\frac{r'_1}{d}} = \floor{\frac{r'_2}{d}} = ... = \floor{\frac{r'_d}{d}}$.

\end{definition}

There are $d^{m-1}$ sets of equivalent RMTs and $d^{m-1}$ sets of sibling RMTs corresponding to the $d^{m-1}$ nodes of de Bruijn graph. We define $Equi_i$ as a set of RMTs which contains RMT $i$ and all of its equivalent RMTs. That is, $Equi_i = \{i, d^{m-1}+i, 2d^{m-1}+i, \cdots, (d-1)d^{m-1}+i \}$, where $0 \leq i \leq d^{m-1}-1$. Similarly, $Sibl_j$ represents a set of sibling RMTs where $Sibl_j = \{d.j, d.j+1, \cdots, d.j+d-1\}$ $(0\leq j \leq d^{m-1}-1)$. That means, the set of sibling RMTs associated to any node $i$ of de Bruijn graph is $Sibl_i$ and the set of equivalent RMTs associated with that node is $Equi_i$. Table~\ref{rln} shows the relationship among the RMTs of $3$-neighborhood $3$-state CAs. An interesting relation is followed in the RMT sequence $(r_i)_{i\in\mathscr{L}}$: if $r_i\in Equi_j$, then $r_{i+1}\in Sibl_j$ ($0\le j\le d^{m-1}-1$).

 \renewcommand{\arraystretch}{0.99}
 \begin{table}[hbtp]
 		\begin{center}
 			\vspace{-1.0em}	
 			\caption{Relations among the RMTs for $3$-neighborhood $3$-state CA}
 			\label{rln}
 			\resizebox{0.9\textwidth}{!}{
 				\begin{tabular}{ccc|ccc}
  \toprule
 	\multicolumn{3}{c|}{\thead{Equivalent Set}} & \multicolumn{3}{c}{\thead{Sibling Set}}\\
 	\thead{\#Set} & \thead{Equivalent RMTs} & \thead{Decimal Equivalent} & \thead{\#Set} & \thead{Sibling RMTs} & \thead{Decimal Equivalent} \\ 
 \midrule
 		
 					$Equi_0$ & 000, 100, 200 & 0, 9, 18 & $Sibl_0$ & ~000, 001, 002 & 0, 1, 2 \\ 
 					$Equi_1$ & 001, 101, 201 & 1, 10, 19 & $Sibl_1$ & ~010, 011, 012 & 3, 4, 5 \\ 
 					$Equi_2$ & 002, 102, 202 & 2, 11, 20 & $Sibl_2$ & ~020, 021, 022 & 6, 7, 8 \\ 
 					$Equi_3$ & 010, 110, 210 & 3, 12, 21 & $Sibl_3$ & ~100, 101, 102 & 9, 10, 11 \\ 
 					$Equi_4$ & 011, 111, 211 & 4, 13, 22 & $Sibl_4$ & ~110, 111, 112 & 12, 13, 14 \\ 
 					$Equi_5$ & 012, 112, 212 & 5, 14, 23 & $Sibl_5$ & ~120, 121, 122 & 15, 16, 17 \\ 
 					$Equi_6$ & 020, 120, 220 & 6, 15, 24 & $Sibl_6$ & ~200, 201, 202 & 18, 19, 20 \\ 
 					$Equi_7$ & 021, 121, 221 & 7, 16, 25 & $Sibl_7$ & ~210, 211, 212 & 21, 22, 23 \\ 
 					$Equi_8$ & 022, 122, 222 & 8, 17, 26 & $Sibl_8$ & ~220, 221, 222 & 24, 25, 26 \\ 
 					\bottomrule
 									\end{tabular}
 								}
 								\end{center}
 								\vspace{-1.0em}	
 								\end{table}

\begin{figure}[hbtp]
\subfloat[$012012120012210102201021102$\label{Chap:reversibility:trans1}]{%
\includegraphics[width=0.47\textwidth, height = 2.7cm]{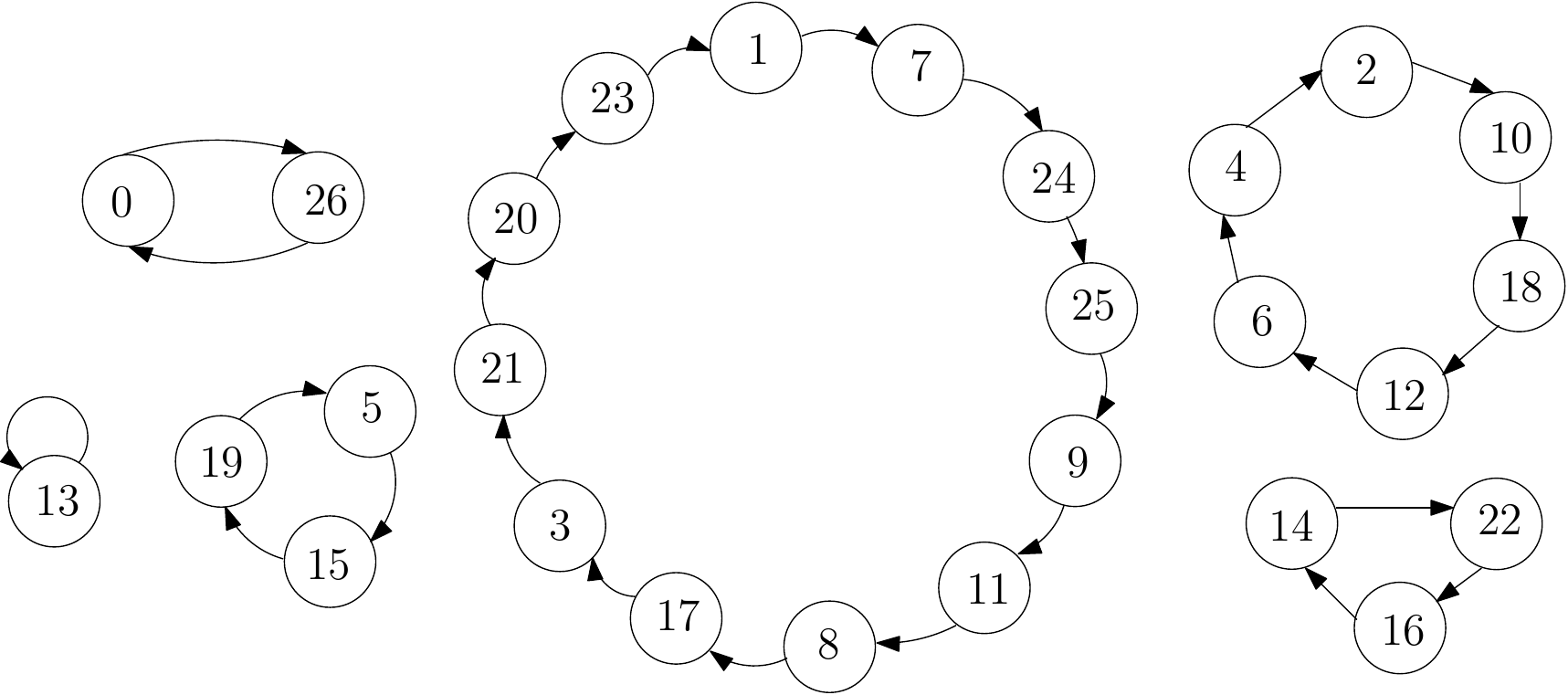}
}
\hfill
\subfloat[$222211112001000000110122221$\label{Chap:reversibility:trans2}]{%
\includegraphics[width=0.47\textwidth, height = 2.7cm]{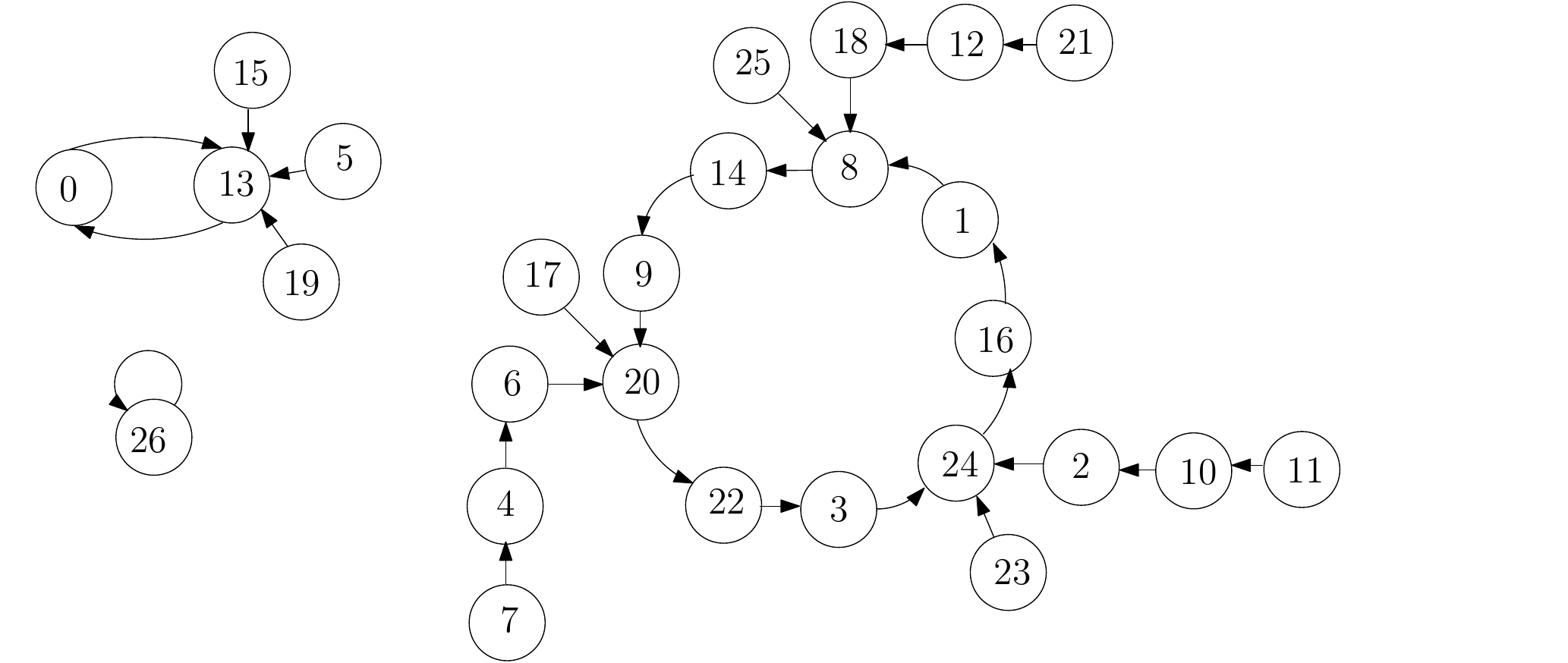}
}
\caption{Configuration transition diagram for two $3$-neighborhood $3$-state CAs with $3$ cells}
\label{fig:config}
			\vspace{-1.5em}
\end{figure}

A configuration is {\emph{reachable}} for a CA, if starting from a distinct initial configuration, that configuration can be reached in the evolution of the CA; otherwise it is {\emph{non-reachable}}.

\begin{definition}\label{Def:reachable}
Let $x \in \mathcal{C}_n$ be a configuration of a CA with global transition function $G_n: \mathcal{C}_n \rightarrow \mathcal{C}_n$. The configuration $x$ is called \textbf{reachable}, if there exists another configuration $y \in \mathcal{C}_n$, such that, $G_n(y)=x$. Otherwise, it is called \textbf{non-reachable}. Here, $\mathcal{C}_n$ is the set of all configurations of length $n$.
\end{definition}

For example, in Figure~\ref{Chap:reversibility:trans1}, all configurations are reachable, because every configuration has a predecessor. However, for Figure~\ref{Chap:reversibility:trans2}, the configurations $5,15,19,21,25,17,7,23$ and $11$ are non-reachable, as they can not be reached from any other configurations. It can be noted that, a CA can be in a non-reachable configuration, only if it is the initial configuration of the CA.

\begin{definition}\label{Def:reversible_finite}
A finite CA of size $n$ with global transition function $G_n: \mathcal{C}_n \rightarrow \mathcal{C}_n$ is \textbf{reversible}, if for each $x \in \mathcal{C}_n$, where $\mathcal{C}_n$ is the set of all configurations of length $n$, there exists exactly one $y \in \mathcal{C}_n$, such that, $G_n(y)=x$. That is, the CA is reversible if $G_n$ is a bijection. Otherwise, the CA is \textbf{irreversible} for that $n$. 
\end{definition}


In Figure~\ref{Chap:reversibility:trans1}, the CA $012012120012210102201021102$ is reversible for $n=3$. Whereas, the CA $222211112001000000110122221$ of Figure~\ref{Chap:reversibility:trans2} is irreversible for $n=3$. For this CA, the configurations $8,20,24$ and $13$ have multiple predecessors.

\section{Reversibility and Semi-reversibility}\label{Chap:semireversible:sec:rev_class}
\noindent For a finite CA of size $n$, reversibility is dependent on $n$. 
A CA is reversible for length $n$, if $G_n$ is bijective, and irreversible, if $G_n$ is not bijective over $\mathcal{C}_n$. Injectivity and surjectivity of $G_n$ are not equivalent to that of $G_P$. Recall that, $G_P$ is the global transition function on the set of all periodic configurations.
If $G_P$ is bijective for periodic configurations of length $n$, for all $n \in \mathbb{N}$, $G_n$ is also bijective for an $n$, and if $G_n$ is not bijective for a size $n$, then $G_P$ is not bijective. However, if $G_n$ is bijective for only some $n$, then, $G_P$ is not a bijection. We are interested in inquiring about injectivity of $G_P$ from injectivity of $G_n$. So, in order to relate $G_n$ with $G_P$, we redefine the notion of reversibility and irreversibility of the CA.

\begin{definition}
\label{chap:semireversibility:def:rev}
A CA with rule $R$ is called \textbf{reversible}, if $G_n$ is bijective on the set of all configurations of length $n$, for each $n\in \mathbb{N}$.
\end{definition}

\begin{definition}
\label{chap:semireversibility:def:irrev}
A CA with rule $R$ is called \textbf{ strictly irreversible}, if $G_n$ is not bijective on the set of all configurations of length $n$, for each $n\in \mathbb{N}$.
\end{definition}

\begin{definition}\label{chap:semireversibility:def:semirev} A CA with rule $R$ is called \textbf{semi-reversible}, if $G_n$ is bijective on the set of all configurations of length $n$, for some $n\in \mathbb{N}$.
\end{definition}

The algorithms by Amoroso and Patt \cite{Amoroso72} as well as of Sutner \cite{suttner91}, identify the reversible CAs of Definition~\ref{chap:semireversibility:def:rev} as reversible for infinite configurations and the strictly irreversible CAs of Definition~\ref{chap:semireversibility:def:irrev} as irreversible for infinite configurations.
These also identify semi-reversible CAs of Definition~\ref{chap:semireversibility:def:semirev}, as irreversible for infinite configurations. Therefore, if a CA with rule $R$ is \emph{reversible}, it is actually reversible for all cases (Case $1$ to $4$), that is, $G_P$, $G$ and $G_F$ as well as $G_n$ for each $n \in \mathbb{N}$ are injective. Similarly, if a CA with local map $R$ is \emph{strictly irreversible}, it is irreversible for all cases, implying each of $G_n$, $G_P$, $G$ and $G_F$ is not injective.

Hence, finite CAs can be classified into three sets -- (1) reversible, (2) strictly irreversible and (3) semi-reversible.
Note that, by the term \emph{reversible}, we mean a CA which is reversible for all sizes $n \in \mathbb{N}$ (Definition~\ref{chap:semireversibility:def:rev}) and by the term \emph{reversible for length $n$}, we mean a CA whose global function $G_n$ is bijective for that length $n$. These two notions are used throughout this paper.

\begin{theorem}\label{Chap:semi-reversibility:Th:strictirreversibility}
A CA is strictly irreversible if and only if it is irreversible for $n=1$.
\end{theorem}

\begin{proof}\begin{description}[leftmargin=1ex]
\item\noindent\underline{\textit{If Part:}} Let us consider that a CA with rule $R$ is irreversible for $n=1$.
That means, for the CA, there exists at least two configurations $x$ and $y$ such that $G_{|x|}(x) = G_{|y|}(y)$. As $|x|=|y|=1$, because of periodic boundary condition, each of these configurations has only a single RMT of the form $(x,x,\cdots,x)$ and $(y,y,\cdots,y)$, where, $x,y \in \{0,1,2,\cdots, d-1\}$. 

Now, let $p$ and $q$ be two configurations of any length using the RMTs $(x,x,\cdots,x)$ and $(y,y,\cdots,y)$ respectively; that is, $p=x^k$ and $q=y^k$ where $k \in \mathbb{N}$. Therefore, $G_{|p|}(p)=G_{|x^k|}(x^k) = (G_{|x|}(x))^k = (G_{|y|}(y))^k = G_{|y^k|}(y^k)= G_{|q|}(q)$. So, for any CA size $n\in \mathbb{N}$, there exists at least two configurations $p$ and $q$ which have same successor. Hence, the CA is irreversible for each $n \in \mathbb{N}$; that is, it is strictly irreversible.

\item\noindent\underline{\textit{Only if Part:}} Assume that a CA is strictly irreversible. Obviously, it is irreversible for $n=1$.
\end{description}\end{proof}

\begin{corollary}\label{Chap:semireversible:corollary:irreversibility}
Let $R$ be the rule of a CA. The CA is strictly irreversible if and only if there exists two RMTs $r$ and $s$, such that $R[r]=R[s]$, where $r=(x,x,\cdots,x)$ and $s=(y,y,\cdots, y)$, $x,y \in S$.
\end{corollary}


\begin{example}
For any $n$-cell ECA, if RMT $0$ and RMT $7$ have same next state values, that is, $R[0]=R[7]$, the CA is strictly irreversible. 
For instance, the ECAs $90(01011010)$ and $30(00011110)$ are strictly irreversible.
\end{example}

Table~\ref{Chap:semireversible:tab:ECA_rev_rules} gives the minimal rules (for the minimal ECA rules and their equivalents, see pages 485--557 of \cite{wolfram86}) for reversible, strictly irreversible and semi-reversible ECAs. There are only $6$ ECA rules, which are reversible according to Definition~\ref{chap:semireversibility:def:rev} and $128$ rules which are strictly irreversible (irreversible for each $n \in \mathbb{N}$). Other $122$ rules are semi-reversible (reversible for some $n \in \mathbb{N}$).
  
\begin{table}[h]
  \setlength{\tabcolsep}{1.5pt}
  \begin{center}
  \caption{The reversible, strictly irreversible and semi-reversible rules of ECAs. Here, only minimal representation \cite{wolfram86} of the rules are written and non-trivial semi-reversible rules are represented in bold face}
  \label{Chap:semireversible:tab:ECA_rev_rules}
  \resizebox{0.9\textwidth}{!}{
 \fbox{
  \begin{tabular}{m{3cm}|m{11.0cm}}
Reversible ECAs & $15$  $51$  $170$  $204$ \\
\hline
\hline
Strictly Irreversible ECAs & $0$ $2$ $4$ $6$ $8$ $10$ $12$ $14$ $18$ $22$ $24$ $26$ $28$ $30$ $32$ $34$ $36$ $38$ $40$ $42$ $44$ $46$ $50$ $54$ $56$ $58$ $60$ $62$ $72$ $74$ $76$ $78$ $90$ $94$ $104$ $106$ $108$ $110$ $122$ $126$ $130$ \\
  \hline
  \hline
   Semi-reversible ECAs & $1$ $3$ $5$  $7$ $9$ $11$ $13$ $19$ $25$ $33$ $35$ $37$ $41$ $73$         $23$ $27$ $29$ $43$ $\mathbf{45}$ $57$ $77$ $\mathbf{105}$ $128$ $132$ $134$ $136$ $138$ $140$  $142$ $146$ $\mathbf{150}$ $152$ $\mathbf{154}$ $156$ $160$ $162$ $164$ $168$ $172$ $178$ $184$ $200$ $232$\\
  \end{tabular}
 } }
  \end{center}
  \vspace{-1.0em}
  \end{table}
 
If a CA is irreversible for $n=1$, it is strictly irreversible (Theorem~\ref{Chap:semi-reversibility:Th:strictirreversibility}). Otherwise, it is either reversible for each $n \in \mathbb{N}$, or semi-reversible. However, among these semi-reversible CAs, some CAs may be present, which are reversible only for a (small) finite set of sizes, and irreversible for an infinite set of sizes. We name such CAs as \textit{trivially} semi-reversible CAs.

\begin{definition}\label{Chap:semireversible:def:triv_semi}
A CA is \textbf{trivially} semi-reversible, if the CA is semi-reversible for a finite set of sizes.
\end{definition}

In Table~\ref{Chap:semireversible:tab:ECA_rev_rules}, the semi-reversible rules written in plain face are trivial semi-revrsible rules. Many of these rules are unbalanced and all of the CAs are not surjective when defined over infinite lattice. However, some CAs exist which are not trivial semi-reversible. For them, an infinite number of sizes exist for which the CA is reversible. In Table~\ref{Chap:semireversible:tab:ECA_rev_rules}, such CAs are marked by bold face. We call these CAs \emph{non-trivial} semi-reversible CAs. 

\begin{definition}\label{Chap:semireversible:def:nontriv_semi}
A semi-reversible CA is \textbf{non-trivial}, if the CA is semi-reversible for an infinite set of sizes.
\end{definition}

%
%

\begin{example}
The ECA rules $45~(00101101)$ and $154~(10011010)$ are reversible for $n =1,3,5,7,\cdots$, that is, when $n$ is odd. The rules $105~(01101001)$ and $150~(10010110)$ are reversible for $n=1,2,4,5,7,8,\cdots $, that is, when $n\ne 3k$, $k\in \mathbb{N}$. Hence, these are \emph{non-trivial} semi-reversible CAs. However, the trivial semi-reversible ECA rules $23~(00010111)$ and $232~(11101000)$ are reversible for $n$ = $1$ and $2$ only, rule $27~(00011011)$ is reversible for $n$ = $1$ and $3$ only and rules $29~(00011101) $, $43~(00101011)$, $57~(00111001)$, $77~(01001101)$, $142~(10001110)$, $156~(10011100)$, $172~(10101100)$, $178~(10110010)$ and $184~(10111000)$ are reversible only for $n$ =$1$, $2$ and $3$.

Let us consider two $4$-neighborhood $2$-state CAs $0101101010101001$ and $1011010011110000$. The first one is trivially semi-reversible, as it is reversible for only $n \in \{1,2\}$, whereas the other one is non-trivially semi-reversible because it is reversible for each odd $n$.
\end{example}


From the discussion of this section, it is evident that, strictly irreversible CAs can be identified beforehand by observing the RMTs $(x,x,\cdots,x)$, $x\in S$. If two such RMTs have same next state value, the CA is strictly irreversible. However, the following questions need to be addressed: 
\begin{enumerate}
\item How can we identify whether a finite CA is reversible for each $n \in \mathbb{N}$ from bijectivity of $G_n$ for some $n$?
\item Can we identify non-trivial semi-reversibility of a CA from a $G_n$ and list the $n$ for which the CA is reversible? These non-trivial semi-reversible CAs are of interest to us. Obviously, for each of these CAs, infinite number of sizes exist for which the CA is reversible. So, our target is to find this infinite set of sizes by some expression(s).
\item How can we identify a trivially semi-reversible CA?
\end{enumerate}
To address these questions, we use a tool, named reachability tree, as discussed in the next section.
%

\section{The Reachability Tree and Reversibility}
\label{Chap:semireversible:sec:rtree}
\noindent In this section, we use a discrete mathematical tool, named \emph{reachability tree}, to analyze the reversibility behavior of any $1$-D finite CA. Reachability tree depicts all reachable configurations of an $n$-cell CA and helps to efficiently decide whether a given $n$-cell CA is reversible or not. It is first introduced for binary CAs in \cite{entcs/DasS09} and then for $d$-state CAs in \cite{jca2015}. 
For an $m$-neighborhood $d$-state CA, the reachability tree can be defined as following --

\begin{definition} \label{chap:semireversibility:def:tree}
Reachability tree of an $n$-cell $m$-neighborhood $d$-state CA is a rooted and edge-labeled $d$-ary tree with $(n+1)$ levels where 
each node $ N_{i.j} ~ (0 \leq i \leq n,~ 0 \leq j \leq d^{i}-1)$ is an ordered list of $d^{m-1}$ sets of RMTs, and the root $N_{0.0}$ is the ordered list of all sets of sibling RMTs. We denote the edges between $N_{i.j} ~ (0 \leq i \leq n-1,~ 0 \leq j \leq d^{i}-1)$ and its possible $d$ children as $E_{i.dj+x} = ( N_{i.j}, N_{i+1.dj+x}, l_{i.dj+x} )$ where $l_{i.dj+x}$ is the label of the edge and $0 \leq x \leq d-1$. Like nodes, the labels are also ordered list of $d^{m-1}$ sets of RMTs. Let us consider that ${\Gamma_{p}}^{N_{i.j}}$ is the $p^{th}$ set of the node $N_{i.j}$, and ${\Gamma_{q}}^{E_{i.dj+x}}$ is the $q^{th}$ set of the label on edge $E_{i.dj+x}$ $(0 \leq p,q \leq d^{m-1} -1)$. So,  $N_{i.j} = ( {\Gamma_{p}}^{N_{i.j}})_{0 \leq p \leq d^{m-1}-1}$ and $l_{i.dj+x} =  ( {\Gamma_{q}}^{E_{i.dj+x}})_{0 \leq q \leq d^{m-1}-1}$. Following are the relations which exist in the tree:

\begin{enumerate}
\item \label{rtd1} [For root] $N_{0.0} = ({\Gamma_{k}^{N_{0.0}}})_{0 \leq k \leq d^{m-1}-1}$, where ${\Gamma_{k}^{N_{0.0}}} = Sibl_k$.

\item \label{rtd2} $\forall r \in {\Gamma_{k}^{N_{i.j}}}, ~ r$ is included in ${\Gamma_{k}^{E_{i.dj +x}}}$, if $R[r] = x, (0 \leq x \leq d-1)$, where $R$ is the rule of the CA. That means, $ {\Gamma_{k}^{N_{i.j}}} = \bigcup\limits_{x} {\Gamma_{k}^{E_{i.dj+x}}}$, $(0 \leq k \leq d^{m-1}-1, 0 \leq i \leq n-1,~ 0 \leq j \leq d^{i}-1)$.

\item \label{rtd4}$\forall r$, if $r \in {\Gamma_{k}^{E_{i.dj+x}}}$, then RMTs of $Sibl_p$, that is $\lbrace d.r \pmod{d^{m}}, d.r+1 \pmod{d^m}, \cdots, d.r+(d-1) \pmod{d^m} \rbrace$ are in ${\Gamma_{k}^{N_{i+1.dj+x}}}$, where $0\leq x \leq d-1$, $0 \leq i \leq n-1$, $0 \leq j \leq d^{i}-1$ and $r \equiv p \pmod{d^{m-1}}$.

\item \label{rtd5} [For level $n-\iota$, $1 \leq \iota \leq m-1$ ] ${\Gamma_{k}^{N_{n-\iota.j}}} = \lbrace y ~|~$ if $ r \in {\Gamma_{k}^{E_{n-\iota-1.j}}} $ then $ y \in \lbrace d.r \pmod{d^m}, d.r+1 \pmod{d^m}, \cdots, d.r+(d-1) \pmod{d^m}\rbrace \cap \lbrace i, i+d^{m-\iota}, i+2d^{m-\iota}, \cdots, i+(d^\iota-1)d^{m-\iota} \rbrace \rbrace$, where $i= \floor{\frac{k}{d^{\iota-1}}}$, $0 \leq k \leq d^{m-1}-1$ and $0 \leq j \leq d^{n-\iota}-1 $.
\end{enumerate}
\end{definition}

Note that, the nodes of level $n-\iota$, $1 \leq \iota \leq m-1$, are different from other intermediate nodes (Points~\ref{rtd5} of Definition~\ref{chap:semireversibility:def:tree}). At level $n-\iota$, $1 \leq \iota \leq m-1$, only some specific RMTs can be present, these RMTs are called \emph{valid} RMTs for this level. In a reachability tree, the root is at level $0$ and the leaves are at level $n$. A sequence of edges from root to leaf, $\langle E_{0.j_1}, E_{1.j_2}, ...,  E_{n-1.j_n}\rangle$, where $0 \leq j_i \leq d^i -1, 1 \leq i \leq n$, constitutes an RMT sequence, where each edge represents a cell's state. If an edge (resp. node) has no RMT, then it is non-reachable edge (resp. node) and represents a non-reachable configuration.

\begin{example}\label{Chap:semireversibility:ex:tree}
The reachability tree for a $5$-cell $4$-neighborhood $2$-state CA with rule $1010101010101010$ is shown in Figure~\ref{Chap:semireversible:fig:rt2}. The details of the nodes and edges for this reachability tree is shown in Table~\ref{Chap:semireversible:Tab:ruleRTree}. This tree contains no non-reachable nodes or edges, so the tree is complete. One reachable configuration $11010$ is represented by the sequence $\langle E_{0.1}, E_{1.3}, E_{2.6}, E_{3.13}, E_{4.26} \rangle$. 

\begin{figure}[!h]
  \centering
    \includegraphics[width= 5.0in, height = 2.5in]{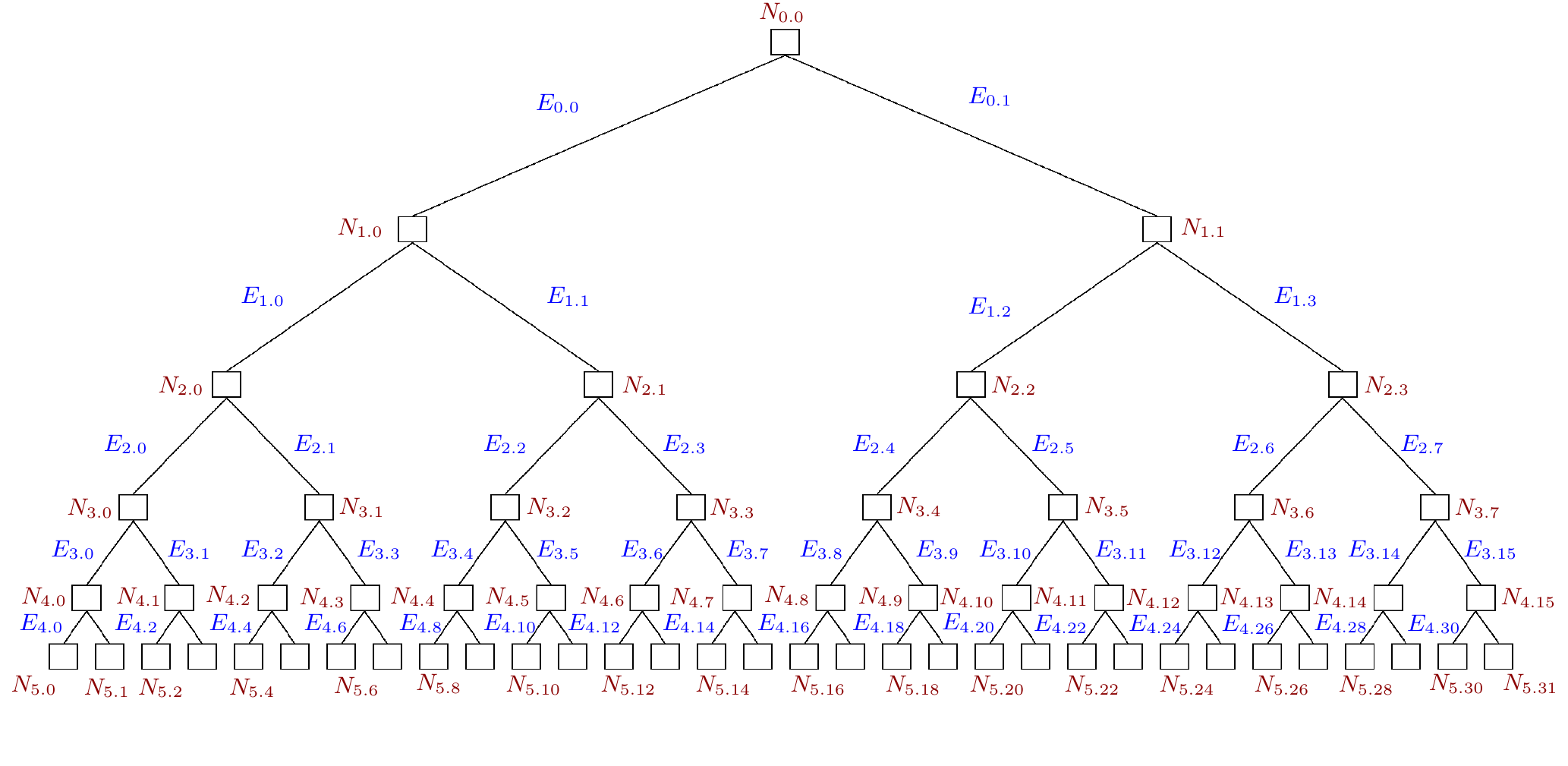} 
   \caption{Reachability tree for $4$-neighborhood $2$-state $5$-cell CA $1010101010101010$}
    \label{Chap:semireversible:fig:rt2}
    \vspace{-2.0em}
\end{figure}
\begin{scriptsize}\setlength\tabcolsep{5pt}
	\centering
\begin{longtable}{|c|p{5.5cm}|p{5cm}|}
\caption{Reachability Tree for CA $4$-neighborhood $2$-state CA $1010101010101010$ (Figure~\ref{Chap:semireversible:fig:rt2}) with $n = 5$}
\label{Chap:semireversible:Tab:ruleRTree}\\
\hline
\textbf{Level} & \textbf{Nodes} & \textbf{Edges} \\
\hline
\endfirsthead
\multicolumn{3}{c}%
{\tablename\ \thetable\ -- \textit{Continued from previous page}} \\
\hline
\textbf{Level} & \textbf{Nodes} & \textbf{Edges} \\
\hline
\endhead
\hline \multicolumn{3}{r}{\textit{Continued on next page}} \\
\endfoot
\hline
\endlastfoot
$0$ & $N_{0.0} = (\{0,1\}$, $\{2,3\}$, $\{4,5\}$, $\{6,7\}$, $\{8,9\}$, $\{10,11\}$, $\{12,13\}$, $\{14,15\})$ & NA \\ 
\hline 
\multirow{2}{*}{$1$} & $N_{1.0}$ = $(\{0,1\}$, $\{4,5\}$, $\{8,9\}$, $\{12,13\}$, $\{0,1\}$, $\{4,5\}$, $\{8,9\}$, $\{12,13\})$ & $E_{0.0}$ = $(\{0\}$, $\{2\}$, $\{4\}$, $\{6\}$, $\{8\}$, $\{10\}$, $\{12\}$, $\{14\})$ \\
 \hhline{~--} 
& $N_{1.1}$ = $(\{2,3\}$, $\{6,7\}$, $\{10,11\}$, $\{14,15\}$, $\{2,3\}$, $\{6,7\}$, $\{10,11\}$, $\{14,15\})$& $E_{0.1}$ = $(\{1\}$, $\{3\}$, $\{5\}$, $\{7\}$, $\{9\}$, $\{11\}$, $\{13\}$, $\{15\})$ \\
\hline
 \multirow{4}{*}{$2$} & $N_{2.0}$ = $(\{0\}$, $\{8\}$, $\{0\}$, $\{8\}$, $\{1\}$, $\{9\}$, $\{1\}$, $\{9\})$ & $E_{1.0}$ = $(\{0\}$, $\{4\}$, $\{8\}$, $\{12\}$, $\{0\}$, $\{4\}$, $\{8\}$, $\{12\})$ \\
  \hhline{~--} 
 & $N_{2.1}$ = $(\{2\}$, $\{10\}$, $\{2\}$, $\{10\}$, $\{3\}$, $\{11\}$, $\{3\}$, $\{11\})$ & $E_{1.1}$ = $(\{1\}$, $\{5\}$, $\{9\}$, $\{13\}$, $\{1\}$, $\{5\}$, $\{9\}$, $\{13\})$ \\
  \hhline{~--} 
 & $N_{2.2}$ = $(\{4\}$, $\{12\}$, $\{4\}$, $\{12\}$, $\{5\}$, $\{13\}$, $\{5\}$, $\{13\})$ & $E_{1.2}$ = $(\{2\}$, $\{6\}$, $\{10\}$, $\{14\}$, $\{2\}$, $\{6\}$, $\{10\}$, $\{14\})$ \\
  \hhline{~--} 
 & $N_{2.3}$ = $(\{6\}$, $\{14\}$, $\{6\}$, $\{14\}$, $\{7\}$, $\{15\}$, $\{7\}$, $\{15\})$ & $E_{1.3}$ = $(\{3\}$, $\{7\}$, $\{11\}$, $\{15\}$, $\{3\}$, $\{7\}$, $\{11\}$, $\{15\})$ \\
 \hline
\multirow{8}{*}{$3$} & $N_{3.0}$ = $(\{0\}$, $\{0\}$, $\{1\}$, $\{1\}$, $\emptyset,\emptyset, \emptyset,\emptyset)$ & $E_{2.0}$ = $(\{0\}$, $\{8\}$, $\{0\}$, $\{8\}$, $\emptyset,\emptyset,\emptyset,\emptyset)$\\
 \hhline{~--} 
& $N_{3.1}$ = $(\emptyset,\emptyset, \emptyset,\emptyset, \{ 2\}$, $\{2\}$, $\{3\}$, $\{3\})$ & $E_{2.1}$ = $(\emptyset,\emptyset, \emptyset,\emptyset$,$\{ 1\}$, $\{9\}$, $\{1\}$, $\{9\})$ \\
 \hhline{~--} 
& $N_{3.2}$ = $(\{4\}$, $\{4\}$, $\{5\}$, $\{5\}$, $\emptyset,\emptyset, \emptyset,\emptyset)$ &  $E_{2.2}$ = $(\{2\}$, $\{10\}$, $\{2\}$, $\{10\}$, $\emptyset,\emptyset, \emptyset,\emptyset)$\\
 \hhline{~--} 
& $N_{3.3}$ = $(\emptyset,\emptyset, \emptyset,\emptyset, \{ 6\}$, $\{6\}$, $\{7\}$, $\{7\})$ & $E_{2.3}$ = $(\emptyset,\emptyset, \emptyset,\emptyset$, $\{3\}$, $\{11\}$, $\{3\}$, $\{11\}$ \\
 \hhline{~--} 
& $N_{3.4}$ = $(\{8\}$, $\{8\}$, $\{9\}$, $\{9\}$, $\emptyset,\emptyset, \emptyset,\emptyset)$ & $E_{2.4}$ = $(\{4\}$, $\{12\}$, $\{4\}$, $\{12\}$, $\emptyset,\emptyset, \emptyset,\emptyset)$\\
 \hhline{~--} 
& $N_{3.5}$ = $(\emptyset,\emptyset, \emptyset,\emptyset, \{ 10\}$, $\{10\}$, $\{11\}$, $\{11\})$ & $N_{3.5}$ = $(\emptyset,\emptyset, \emptyset,\emptyset, \{5\}$, $\{13\}$, $\{5\}$, $\{13\})$ \\
 \hhline{~--} 
& $N_{3.6}$ = $(\{12\}$, $\{12\}$, $\{13\}$, $\{13\}$, $\emptyset,\emptyset, \emptyset,\emptyset)$ & $E_{2.6}$ = $(\{6\}$, $\{14\}$, $\{6\}$, $\{14\}$, $\emptyset,\emptyset, \emptyset,\emptyset)$\\
 \hhline{~--} 
& $N_{3.7}$ = $(\emptyset,\emptyset, \emptyset,\emptyset, \{ 14\}$, $\{14\}$, $\{15\}$, $\{15\})$ & $E_{2.7}$ = $(\emptyset,\emptyset, \emptyset,\emptyset,\{7\}$, $\{15\}$, $\{7\}$, $\{15\})$ \\
\hline
\multirow{16}{*}{$4$} & $N_{4.0}$ = $(\{0\}$, $\{1\}$, $\emptyset,\emptyset, \emptyset,\emptyset, \emptyset,\emptyset)$ &  $E_{3.0}$ = $(\{0\}$, $\{0\}$, $\emptyset,\emptyset, \emptyset,\emptyset, \emptyset,\emptyset)$\\
 \hhline{~--} 
& $N_{4.1}$ = $(\emptyset,\emptyset, \{ 2\}$, $\{3\}$, $\emptyset,\emptyset, \emptyset,\emptyset)$ & $E_{3.1}$ = $(\emptyset,\emptyset, \{1\}$, $\{1\}$, $\emptyset,\emptyset, \emptyset,\emptyset)$\\
 \hhline{~--} 
& $N_{4.2}$ = $(\emptyset,\emptyset, \emptyset,\emptyset, \{ 4\}$, $\{5\}$, $\emptyset,\emptyset)$ & $E_{3.2}$ = $(\emptyset,\emptyset, \emptyset,\emptyset, \{ 2\}$, $\{2\}$, $\emptyset,\emptyset)$\\
 \hhline{~--} 
& $N_{4.3}$ = $(\emptyset,\emptyset, \emptyset,\emptyset, \emptyset,\emptyset, \{ 6\}$, $\{7\})$ & $E_{3.3}$ = $(\emptyset,\emptyset, \emptyset,\emptyset, \emptyset,\emptyset, \{ 3\}$, $\{3\})$ \\
 \hhline{~--} 
& $N_{4.4}$ = $(\{8\}$, $\{9\}$, $\emptyset,\emptyset, \emptyset,\emptyset, \emptyset,\emptyset)$ &  $E_{3.4}$ = $(\{4\}$, $\{4\}$, $\emptyset,\emptyset, \emptyset,\emptyset, \emptyset,\emptyset)$\\
 \hhline{~--} 
& $N_{4.5}$ = $(\emptyset,\emptyset, \{ 10\}$, $\{11\}$, $\emptyset,\emptyset, \emptyset,\emptyset)$ & $E_{3.5}$ = $(\emptyset,\emptyset, \{ 5\}$, $\{5\}$, $\emptyset,\emptyset, \emptyset,\emptyset)$\\
 \hhline{~--} 
& $N_{4.6}$ = $(\emptyset,\emptyset, \emptyset,\emptyset, \{ 12\}$, $\{13\}$, $\emptyset,\emptyset)$ & $E_{3.6}$ = $(\emptyset,\emptyset, \emptyset,\emptyset, \{ 6\}$, $\{6\}$, $\emptyset,\emptyset)$ \\
 \hhline{~--} 
& $N_{4.7}$ = $(\emptyset,\emptyset, \emptyset,\emptyset, \emptyset,\emptyset, \{ 14\}$, $\{15\})$ & $E_{3.7}$ = $(\emptyset,\emptyset, \emptyset,\emptyset, \emptyset,\emptyset, \{ 7\}$, $\{7\})$ \\
 \hhline{~--} 
& $N_{4.8}$ = $(\{0\}$, $\{1\}$, $\emptyset,\emptyset, \emptyset,\emptyset, \emptyset,\emptyset)$ & $E_{3.8}$ = $(\{8\}$, $\{8\}$, $\emptyset,\emptyset, \emptyset,\emptyset, \emptyset,\emptyset)$\\
 \hhline{~--} 
& $N_{4.9}$ = $(\emptyset,\emptyset, \{ 2\}$, $\{3\}$, $\emptyset,\emptyset, \emptyset,\emptyset)$ & $E_{3.9}$ = $(\emptyset,\emptyset, \{ 9\}$, $\{9\}$, $\emptyset,\emptyset, \emptyset,\emptyset)$\\
 \hhline{~--} 
& $N_{4.10}$ = $(\emptyset,\emptyset, \emptyset,\emptyset, \{ 4\}$, $\{5\}$, $\emptyset,\emptyset)$ & $E_{3.10}$ = $(\emptyset,\emptyset, \emptyset,\emptyset, \{ 10\}$, $\{10\}$, $\emptyset,\emptyset)$\\
 \hhline{~--} 
& $N_{4.11}$ = $(\emptyset,\emptyset, \emptyset,\emptyset, \emptyset,\emptyset, \{ 6\}$, $\{7\})$ & $E_{3.11}$ = $(\emptyset,\emptyset, \emptyset,\emptyset, \emptyset,\emptyset, \{ 11\}$, $\{11\})$ \\
 \hhline{~--} 
& $N_{4.12}$ = $(\{8\}$, $\{9\}$, $\emptyset,\emptyset, \emptyset,\emptyset, \emptyset,\emptyset)$ & $E_{3.12}$ = $(\{12\}$, $\{12\}$, $\emptyset,\emptyset, \emptyset,\emptyset, \emptyset,\emptyset)$\\
 \hhline{~--} 
& $N_{4.13}$ = $(\emptyset,\emptyset, \{ 10\}$, $\{11\}$, $\emptyset,\emptyset, \emptyset,\emptyset)$ & $E_{3.13}$ = $(\emptyset,\emptyset, \{ 13\}$, $\{13\}$, $\emptyset,\emptyset, \emptyset,\emptyset)$ \\
 \hhline{~--} 
& $N_{4.14}$ = $(\emptyset,\emptyset, \emptyset,\emptyset, \{ 12\}$, $\{13\}$, $\emptyset,\emptyset)$ & $E_{3.14}$ = $(\emptyset,\emptyset, \emptyset,\emptyset, \{ 14\}$, $\{14\}$, $\emptyset,\emptyset)$\\
 \hhline{~--} 
& $N_{4.15}$ = $(\emptyset,\emptyset, \emptyset,\emptyset, \emptyset,\emptyset, \{ 14\}$, $\{15\})$ & $E_{3.15}$ = $(\emptyset,\emptyset, \emptyset,\emptyset, \emptyset,\emptyset, \{ 15\}$, $\{15\})$ \\
\hline

\multirow{32}{*}{$5$} & $N_{5.0}$ = $(\{0,1\}$, $\emptyset$, $\emptyset,\emptyset, \emptyset,\emptyset, \emptyset,\emptyset)$ & $E_{4.0}$ = $(\{0\}$, $\emptyset$, $\emptyset,\emptyset, \emptyset,\emptyset, \emptyset,\emptyset)$ \\
 \hhline{~--} 
& $N_{5.1}$ = $(\emptyset$, $\{2,3\}$, $\emptyset,\emptyset, \emptyset,\emptyset, \emptyset,\emptyset)$ & $E_{4.1}$ = $(\emptyset$, $\{1\}$, $\emptyset,\emptyset, \emptyset,\emptyset, \emptyset,\emptyset)$ \\
 \hhline{~--} 
& $N_{5.2}$ = $(\emptyset,\emptyset, \{ 4,5\}$, $\emptyset$, $\emptyset,\emptyset, \emptyset,\emptyset)$ & $E_{4.2}$ = $(\emptyset,\emptyset, \{ 2\}$, $\emptyset$, $\emptyset,\emptyset, \emptyset,\emptyset)$ \\
 \hhline{~--} 
& $N_{5.3}$ = $(\emptyset,\emptyset, \emptyset$, $\{6,7\}$, $\emptyset,\emptyset, \emptyset,\emptyset)$ & $E_{4.3}$ = $(\emptyset,\emptyset, \emptyset$, $\{3\}$, $\emptyset,\emptyset, \emptyset,\emptyset)$\\
 \hhline{~--} 
& $N_{5.4}$= $(\emptyset,\emptyset, \emptyset,\emptyset, \{ 8,9\}$, $\emptyset$, $\emptyset,\emptyset)$ & $E_{4.4}$= $(\emptyset,\emptyset, \emptyset,\emptyset, \{ 4\}$, $\emptyset$, $\emptyset,\emptyset)$\\
 \hhline{~--} 
& $N_{5.5}$ = $(\emptyset,\emptyset, \emptyset,\emptyset, \emptyset$, $\{10,11\}$, $\emptyset,\emptyset)$ & $E_{4.5}$ = $(\emptyset,\emptyset, \emptyset,\emptyset, \emptyset$, $\{5\}$, $\emptyset,\emptyset)$\\
 \hhline{~--} 
& $N_{5.6}$ =$(\emptyset,\emptyset, \emptyset,\emptyset, \emptyset,\emptyset, \{ 12,13\}$, $\emptyset)$ & $E_{4.6}$ =$(\emptyset,\emptyset, \emptyset,\emptyset, \emptyset,\emptyset, \{6\}$, $\emptyset)$\\
 \hhline{~--} 
& $N_{5.7}$ =$(\emptyset,\emptyset, \emptyset,\emptyset, \emptyset,\emptyset, \emptyset$, $\{14,15\})$ & $E_{4.7}$ =$(\emptyset,\emptyset, \emptyset,\emptyset, \emptyset,\emptyset, \emptyset$, $\{7\})$\\
\hhline{~--} 
& $N_{5.8}$ = $(\{0,1\}$, $\emptyset$, $\emptyset,\emptyset, \emptyset,\emptyset, \emptyset,\emptyset)$ & $E_{4.8}$ = $(\{8\}$, $\emptyset$, $\emptyset,\emptyset, \emptyset,\emptyset, \emptyset,\emptyset)$  \\
 \hhline{~--} 
& $N_{5.9}$ = $(\emptyset$, $\{2,3\}$, $\emptyset,\emptyset, \emptyset,\emptyset, \emptyset,\emptyset)$ & $E_{4.9}$ = $(\emptyset$, $\{9\}$, $\emptyset,\emptyset, \emptyset,\emptyset, \emptyset,\emptyset)$ \\
 \hhline{~--} 
& $N_{5.10}$ = $(\emptyset,\emptyset, \{ 4,5\}$, $\emptyset$, $\emptyset,\emptyset, \emptyset,\emptyset)$ & $E_{4.10}$ = $(\emptyset,\emptyset, \{10\}$, $\emptyset$, $\emptyset,\emptyset, \emptyset,\emptyset)$\\
 \hhline{~--} 
& $N_{5.11}$ = $(\emptyset,\emptyset, \emptyset$, $\{6,7\}$, $\emptyset,\emptyset, \emptyset,\emptyset)$ & $E_{4.11}$ = $(\emptyset,\emptyset, \emptyset$, $\{11\}$, $\emptyset,\emptyset, \emptyset,\emptyset)$ \\
 \hhline{~--} 
& $N_{5.12}$= $(\emptyset,\emptyset, \emptyset,\emptyset, \{ 8,9\}$, $\emptyset$, $\emptyset,\emptyset)$ & $E_{4.12}$= $(\emptyset,\emptyset, \emptyset,\emptyset, \{ 12\}$, $\emptyset$, $\emptyset,\emptyset)$\\
 \hhline{~--} 
& $N_{5.13}$ = $(\emptyset,\emptyset, \emptyset,\emptyset, \emptyset$, $\{10,11\}$, $\emptyset,\emptyset)$ & $E_{4.13}$ = $(\emptyset,\emptyset, \emptyset,\emptyset, \emptyset$, $\{13\}$, $\emptyset,\emptyset)$\\
 \hhline{~--} 
& $N_{5.14}$ =$(\emptyset,\emptyset, \emptyset,\emptyset, \emptyset,\emptyset, \{ 12,13\}$, $\emptyset)$ & $E_{4.14}$ =$(\emptyset,\emptyset, \emptyset,\emptyset, \emptyset,\emptyset, \{14\}$, $\emptyset)$\\
 \hhline{~--} 
& $N_{5.15}$ = =$(\emptyset,\emptyset, \emptyset,\emptyset, \emptyset,\emptyset, \emptyset$, $\{14,15\})$ & $E_{4.15}$ =$(\emptyset,\emptyset, \emptyset,\emptyset, \emptyset,\emptyset, \emptyset$, $\{15\})$\\
 \hhline{~--} 
& $N_{5.16}$ = $(\{0,1\}$, $\emptyset$, $\emptyset,\emptyset, \emptyset,\emptyset, \emptyset,\emptyset)$ & $E_{4.16}$ = $(\{0\}$, $\emptyset$, $\emptyset,\emptyset, \emptyset,\emptyset, \emptyset,\emptyset)$  \\
 \hhline{~--} 
& $N_{5.17}$ = $(\emptyset$, $\{2,3\}$, $\emptyset,\emptyset, \emptyset,\emptyset, \emptyset,\emptyset)$ & $E_{4.17}$ = $(\emptyset$, $\{1\}$, $\emptyset,\emptyset, \emptyset,\emptyset, \emptyset,\emptyset)$ \\
 \hhline{~--} 
& $N_{5.18}$ = $(\emptyset,\emptyset, \{ 4,5\}$, $\emptyset$, $\emptyset,\emptyset, \emptyset,\emptyset)$ & $E_{4.18}$ = $(\emptyset,\emptyset, \{ 2\}$, $\emptyset$, $\emptyset,\emptyset, \emptyset,\emptyset)$\\
& $N_{5.19}$ = $(\emptyset,\emptyset, \emptyset$, $\{6,7\}$, $\emptyset,\emptyset, \emptyset,\emptyset)$ & $E_{4.19}$ = $(\emptyset,\emptyset, \emptyset$, $\{3\}$, $\emptyset,\emptyset, \emptyset,\emptyset)$\\
 \hhline{~--} 
& $N_{5.20}$= $(\emptyset,\emptyset, \emptyset,\emptyset, \{ 8,9\}$, $\emptyset$, $\emptyset,\emptyset)$ & $E_{4.20}$= $(\emptyset,\emptyset, \emptyset,\emptyset, \{ 4\}$, $\emptyset$, $\emptyset,\emptyset)$\\
 \hhline{~--} 
& $N_{5.21}$ = $(\emptyset,\emptyset, \emptyset,\emptyset, \emptyset$, $\{10,11\}$, $\emptyset,\emptyset)$ & $E_{4.21}$ = $(\emptyset,\emptyset, \emptyset,\emptyset, \emptyset$, $\{5\}$, $\emptyset,\emptyset)$\\
 \hhline{~--} 
& $N_{5.22}$ =$(\emptyset,\emptyset, \emptyset,\emptyset, \emptyset,\emptyset, \{ 12,13\}$, $\emptyset)$ & $E_{4.22}$ =$(\emptyset,\emptyset, \emptyset,\emptyset, \emptyset,\emptyset, \{6\}$, $\emptyset)$\\
 \hhline{~--} 
& $N_{5.23}$ = $(\emptyset,\emptyset, \emptyset,\emptyset, \emptyset,\emptyset, \emptyset$, $\{14,15\})$ & $E_{4.23}$ =$(\emptyset,\emptyset, \emptyset,\emptyset, \emptyset,\emptyset, \emptyset$, $\{7\})$\\
 \hhline{~--} 
& $N_{5.24}$ = $(\{0,1\}$, $\emptyset$, $\emptyset,\emptyset, \emptyset,\emptyset, \emptyset,\emptyset)$ & $E_{4.24}$ = $(\{8\}$, $\emptyset$, $\emptyset,\emptyset, \emptyset,\emptyset, \emptyset,\emptyset)$  \\
 \hhline{~--} 
& $N_{5.25}$ = $(\emptyset$, $\{2,3\}$, $\emptyset,\emptyset, \emptyset,\emptyset, \emptyset,\emptyset)$ & $E_{4.25}$ = $(\emptyset$, $\{9\}$, $\emptyset,\emptyset, \emptyset,\emptyset, \emptyset,\emptyset)$\\
 \hhline{~--} 
& $N_{5.26}$ = $(\emptyset,\emptyset, \{ 4,5\}$, $\emptyset$, $\emptyset,\emptyset, \emptyset,\emptyset)$ & $E_{4.26}$ = $(\emptyset,\emptyset, \{ 10\}$, $\emptyset$, $\emptyset,\emptyset, \emptyset,\emptyset)$\\
 \hhline{~--} 
& $N_{5.27}$ = $(\emptyset,\emptyset, \emptyset$, $\{6,7\}$, $\emptyset,\emptyset, \emptyset,\emptyset)$ & $E_{4.27}$ = $(\emptyset,\emptyset, \emptyset$, $\{11\}$, $\emptyset,\emptyset, \emptyset,\emptyset)$\\
 \hhline{~--} 
& $N_{5.28}$= $(\emptyset,\emptyset, \emptyset,\emptyset, \{ 8,9\}$, $\emptyset$, $\emptyset,\emptyset)$ & $E_{4.28}$= $(\emptyset,\emptyset, \emptyset,\emptyset, \{ 12\}$, $\emptyset$, $\emptyset,\emptyset)$\\
 \hhline{~--} 
& $N_{5.29}$ = $(\emptyset,\emptyset, \emptyset,\emptyset, \emptyset$, $\{10,11\}$, $\emptyset,\emptyset)$ & $E_{4.29}$ = $(\emptyset,\emptyset, \emptyset,\emptyset, \emptyset$, $\{13\}$, $\emptyset,\emptyset)$\\
 \hhline{~--} 
& $N_{5.30}$ =$(\emptyset,\emptyset, \emptyset,\emptyset, \emptyset,\emptyset, \{ 12,13\}$, $\emptyset)$ & $E_{4.30}$ =$(\emptyset,\emptyset, \emptyset,\emptyset, \emptyset,\emptyset, \{14\}$, $\emptyset)$\\
 \hhline{~--} 
& $N_{5.31}$ = $(\emptyset,\emptyset, \emptyset,\emptyset, \emptyset,\emptyset, \emptyset$, $\{14,15\})$ & $E_{4.31}$ =$(\emptyset,\emptyset, \emptyset,\emptyset, \emptyset,\emptyset, \emptyset$, $\{15\})$\\
\end{longtable}
\end{scriptsize}
\end{example}

In a reachability tree, some nodes are \emph{balanced}, some are not. 
\begin{definition}
\label{def:balancednode}
A node is called \textbf{balanced} if it has equal number of RMTs corresponding to each of the $d$-states possible; otherwise it is \textbf{unbalanced} \cite{jca2015}.

\end{definition}
For instance, all nodes of the balanced rule (see Definition~\ref{Def:balancedrule}) $1010101010101010$ of Example~\ref{Chap:semireversibility:ex:tree} are balanced.

%

\subsection{Reversibility Theorems for Reachability Tree}
\noindent Here, we explore reachability tree to report the conditions it needs to fulfill to represent a reversible CA of size $n$. See Appendix for proofs of these theorems.
\begin{theorem}
\label{chap:semireversibility:th:revth1}
The reachability tree of a finite reversible CA of length $n$ $(n \geq m)$ is complete.
\end{theorem} 


\begin{theorem}
\label{chap:semireversibility:th:revth2}
The reachability tree of a finite CA of length $n$ $ (n \geq m)$ is complete if and only if
\begin{enumerate}[topsep=0pt,itemsep=0ex,partopsep=2ex,parsep=1ex]
\item \label{c2}  The label $l_{n-\iota.j}$, for any $j$, contains only $d^{\iota-1}$ RMTs, where $1 \leq \iota \leq m-1$; that is, \[\mid \bigcup\limits_{0 \leq k \leq d^{m-1} -1} {\Gamma_{k}^{E_{n-\iota.j}}}\mid = d^{\iota-1}\]

\item \label{c3} Each other label $l_{i.j}$ contains $d^{m-1}$ RMTs, where $ 0 \leq i \leq n-m$; that is, \[ \mid \bigcup\limits_{0 \leq k \leq d^{m-1} -1} {\Gamma_{k}}^{E_{i.j}}\mid = d^{m-1}\]
\end{enumerate}
\end{theorem}

\begin{corollary}
\label{chap:semireversibility:th:revcor1}
The nodes of a reachability tree of a reversible CA of length $n$ $(n \geq m)$ contains
\begin{enumerate}[topsep=0pt,itemsep=0ex,partopsep=2ex,parsep=1ex]
\item $d$ RMTs, if the node is in level $n$, i.e. $ \mid \bigcup\limits_{0 \leq k \leq d^{m-1} -1}{\Gamma_{k}^{N_{n.j}}} \mid = d$ for any $j$.

\item $d^\iota $ RMTs, if the node is at level $n-\iota$ i.e, $ \mid \bigcup\limits_{0 \leq k \leq d^{m-1} -1}{\Gamma_{k}^{N_{n-\iota.j}}} \mid = d^\iota$ for any $j$, where $1 \leq \iota \leq m-1$.

\item $d^m$ RMTs for all other nodes $N_{i.j}$, $ \mid \bigcup\limits_{0 \leq k \leq d^{m-1} -1}{\Gamma_{k}^{N_{i.j}}} \mid = d^m$ where ${ 0 \leq i \leq n-m}$.
\end{enumerate}
\end{corollary}


\begin{corollary} 
\label{chap:semireversibility:th:revcor2}
The nodes of the reachability tree of an $n$-cell $(n \geq m)$ reversible CA are balanced.
\end{corollary}


\begin{corollary}
\label{chap:semireversibility:th:revcor3}
Let $R$ be the rule of a CA. Now, the CA is trivially semi-reversible (irreversible for each $n\geq m$) if the following conditions are satisfied--
\begin{enumerate}[topsep=0pt,itemsep=0ex,partopsep=2ex,parsep=1ex]
\item $R$ is unbalanced and
\item no two RMTs $r$ and $s$ exist, such that $r=(x,x\cdots,x)$ and $s=(y,y,\cdots,y)$, $x,y \in S$ and $R[r]=R[s]$.
\end{enumerate}
\end{corollary}

%
%

Reachability tree helps to identify whether a CA is reversible for a size $n$. If the reachability tree of an $n$-cell CA has all possible edges/nodes, then this implies that, all configurations are reachable for the CA. That is, if the tree is complete, the CA is reversible for that cell length $n$ (see Theorem~\ref{chap:semireversibility:th:revth1}).
However, for a CA to be reversible for an $n$, $(n \geq m)$, the reachability tree needs to satisfy the following two conditions --
\begin{enumerate} 
\item The nodes of the reachability tree are to be balanced.
\item Each node of level $n$ is to have $d$ RMTs, each node of level $n-\iota$, $1 \leq \iota \leq m-1$, is to contain $d^\iota $ RMTs, whereas all other nodes are to be made with $d^m$ RMTs.
\end{enumerate}

For instance, the tree of Figure~\ref{Chap:semireversible:fig:rt2} is complete. Further, each node and edge of the reachability tree satisfies the above reversibility conditions. So, by Theorem~\ref{chap:semireversibility:th:revth1} and Theorem~\ref{chap:semireversibility:th:revth2}, the CA is reversible for $n=5$. 
\subsection{Minimized Reachability Tree}\label{Chap:semireversible:sec:tree-cons}
\noindent However, for very large $n$, the tree grows exponentially with large number of nodes, so it becomes impossible to deal with. Nevertheless, one can observe that, in a reachability tree, some nodes are similar to each other. For example,

\begin{enumerate} 
\item if $N_{i.j} = N_{i.k}$ when $j \neq k$ for any $i$, then both the nodes are roots of two similar sub-trees. So, we can proceed with only one node. Similarly, if $l_{i.j} = l_{i.k} ~(j \neq k)$, then also we can proceed with only one edge. 
\item if $N_{i.j} = N_{i'.k}$ when $i>i', 0\leq i,i' \leq n-\iota-1$ and $1 \leq \iota \leq m-1$, then the nodes that follow $N_{i'.k}$ are similar with the nodes followed by $N_{i.j}$. Therefore, we need not to explicitly develop the sub-tree of $N_{i.j}$. 
\end{enumerate}

It is observed that after few levels, no unique node is generated. So, for arbitrary large $n$, we need not to develop the whole tree, rather, the \emph{minimized reachability tree} \cite{jca2015} is developed which stores only the \emph{unique} nodes generated in the tree.
Here, the steps of forming the minimized tree (consisting of all possible unique nodes for the CA) are enlisted briefly.
\begin{description} 
\item[Step 1:]\label{minimizedTree:step_1} Form the root $N_{0.0}$ of the tree. Set index of level, $i\leftarrow 0$ and set of levels of node $N_{0.0} \leftarrow \{0\}$. As root is always unique, move to the next level.

\item[Step 2:]\label{minimizedTree:step_2} For each node $\mathcal{N}$ of level $i$, 
\begin{enumerate}[leftmargin=0pt,topsep=0pt,itemsep=0ex,partopsep=2ex,parsep=1ex]
\item Find the children of $\mathcal{N}$ for level $i+1$

\item For each child node, say, $N_{i+1.j}$, check whether it is unique or not --
\begin{enumerate}[topsep=0pt,itemsep=0ex,partopsep=2ex,parsep=1ex]
\item If $N_{i+1.j} = N_{i+1.k}$ when $j > k$, then both the nodes are roots of two similar sub-trees. So, we discard $N_{i+1.j}$ and add a link from the parent node $\mathcal{N}$ to $N_{i+1.k}$.

\item If $N_{i+1.j} = N_{i'.k}$ when $i+1>i'$, 
then the nodes that follow $N_{i'.k}$ are similar with the nodes followed by $N_{i+1.j}$. Therefore, we discard $N_{i+1.j}$ and add a link from $\mathcal{N}$ to $N_{i'.k}$. The level of node $N_{i'.k}$ is updated by adding $i+1$ with the existing set of levels. For example, say, $N_{i'.k}$ was present only in level $i'$, then the previous set of levels of $N_{i'.k}$ is $\{i'\}$. This set is now updated as $\{i',i+1\}$. In this case, a \emph{loop} is embodied between levels $i+1$ and $i'$. Presence of this loop is indicated by this set of levels of node $N_{i'.k}$.

\item Otherwise, the node $N_{i+1.j}$ is unique. Add this node in the minimized reachability tree and update its set of levels according to the set of levels of the parent node $\mathcal{N}$.
\end{enumerate}

\item Whenever, a new level is added to the set of levels of an existing node $\mathcal{N}'$ of the minimized tree, the set of levels for each node of the subtree of $\mathcal{N}'$ are updated according to the new set of levels of $\mathcal{N}'$. 
\end{enumerate} 
\item[Step 3:]\label{minimizedTree:step_3} If a new node is added in the tree, update $i \leftarrow i+1$ and go to Step~$2$.

\item[Step 4:]\label{minimizedTree:step_4} If no new node is added in the tree, construction of the minimized reachability tree is complete.
\end{description}



Strictly speaking, the minimized reachability tree is not a tree; rather, it is a directed graph, and the directions are necessary to reconstruct the original tree. In our further discussion, however, we call it as tree. Nevertheless, if $n_0$ is the level when the last unique node is added in the minimized reachability tree corresponding to a CA, then, for any $n\ge n_0$, the minimized tree for the CA does not change. However, if $n<n_0$, then we need to develop the minimized reachability tree up to level $n-m$ and generate nodes of $n-\iota$, $1 \leq \iota \leq m-1$ according to Point~\ref{rtd4} of Definition~\ref{chap:semireversibility:def:tree}.


In the minimized tree, if a node $\mathcal{N}$ is generated in level $i$ as well as level $i'$, $i'\geq i$, then the unique node of level $i$ gets two node levels $\lbrace i, i'\rbrace$ and a loop of length $(i'-i)$ is associated with the node. This loop implies that the node reappears at levels on the arithmetic series $i + (i - i')$, $i + 2(i-i')$, $i + 3(i-i')$, etc. Moreover, a node $\mathcal{N}$ can be part of more than one loop, which implies that, $\mathcal{N}$ can appear at levels implied by each of the loops. However, if we observe in more detail, we can find that, although every loop confirms presence of $\mathcal{N}$, but all loops are not significant in the tree.
We need to keep only the relevant loops in the tree and discard the other loops.



\begin{example} Consider the minimized reachability tree of ECA $75 (01001011)$ shown in Figure~\ref{fig:rt3}. Details of nodes of the minimized tree along with the updated levels associated with each node is shown in Table~\ref{tab:ex:ECA75}. The minimized tree has only $21$ nodes and the height of the tree is $5$. Therefore, for each $n\ge 5$, the minimized tree remains the same and conveys all information regarding reversibility of the CA for that $n$.
\begin{figure}[!h]
		 \centering
			\includegraphics[width= 4.5in, height = 3.1in]{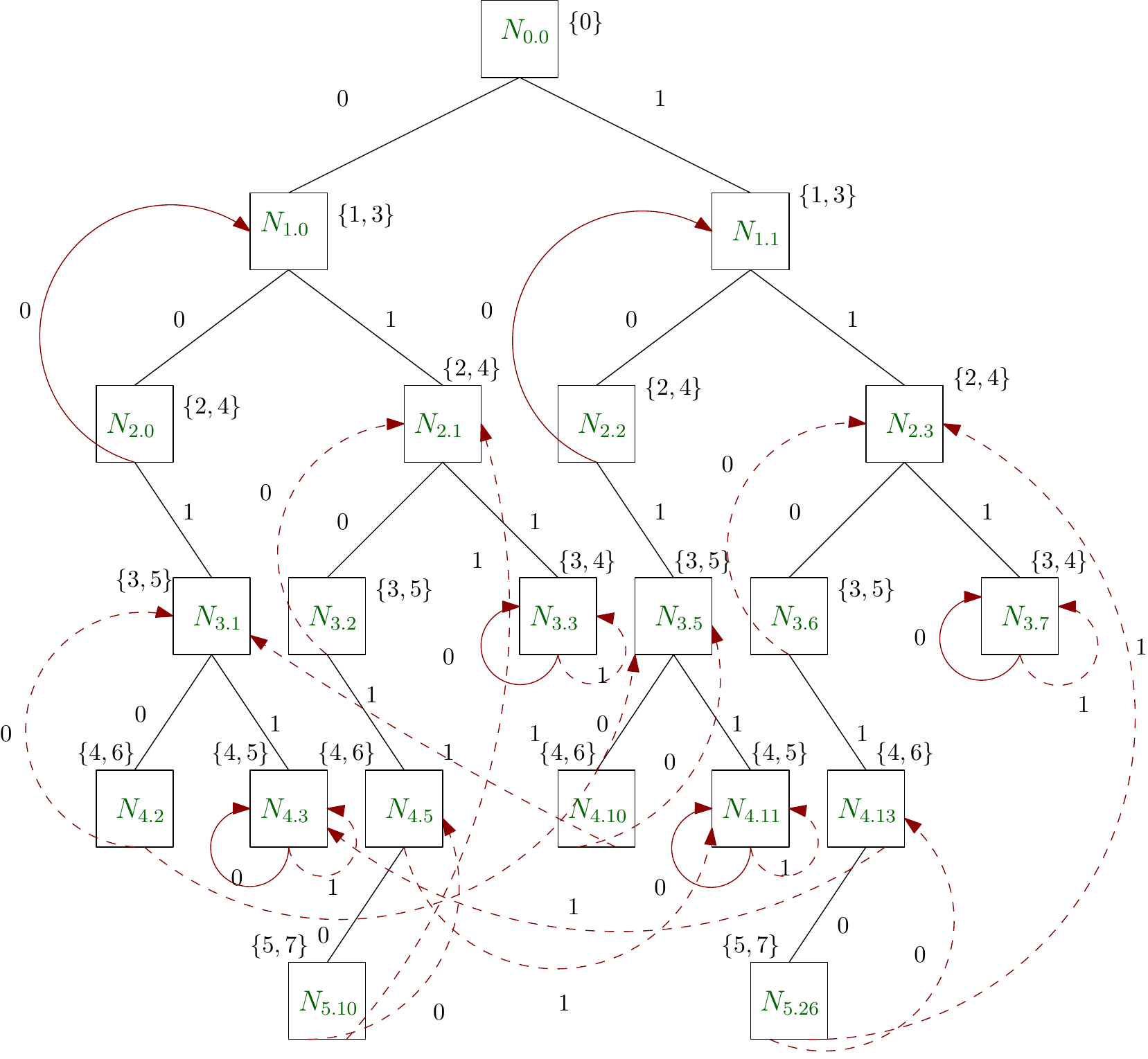}
			\caption{Minimized reachability tree of ECA $75 (01001011)$}\label{fig:rt3}
			\vspace{-1.5em}
\end{figure}
Here, most of the nodes of the minimized tree has multiple levels associated with it reflecting the loops. In Figure~\ref{fig:rt3}, the irrelevant loops in the tree are shown in dashed line. 
\begin{table}[!h]
\renewcommand{\arraystretch}{1.1}
\centering
\caption{Details of minimized tree for ECA rule $75 (01001011)$}
\vspace{-0.5em}
\label{tab:ex:ECA75}
\resizebox{1.0\textwidth}{!}{
\vspace{-\topsep} 
\begin{tabular}{|c|c|c|c|c|}
\hline 
$i$ &  Unique Node & Level of the Node & \multicolumn{1}{c|}{\begin{tabular}{c}Unbalanced\\ for level?\end{tabular}}&
	\multicolumn{1}{c|}{ \begin{tabular}{c} Irreversibility \\ Expression \end{tabular}}\\ 
\hline 
$0$ & $N_{0.0} = (\{0,1\}$, $\{2,3\}$, $\{4,5\}$, $\{6,7\})$ & $\lbrace 0 \rbrace$ & NA & NA\\ 
\hline 

\multirow{2}{*}{$1$} & $N_{3.0} \equiv N_{1.0} =(\emptyset, \{4,5\}$, $\{0,1,2,3\}$, $\{6,7\})$ & $\lbrace1, 3\rbrace$ & $n-1$ & $n = 2j+2$\\  
\hhline{~----} 
& $N_{3.4} \equiv N_{1.1} = (\{0,1,2,3\}$, $\{6,7\}, \emptyset, \{4,5\})$ & $\lbrace1, 3\rbrace$ & $n-1$ & $n=2j+2$\\  
\hline 

\multirow{4}{*}{$2$} & $N_{2.0} = (\emptyset, \{0,1,2,3\}$, $\{4,5\}$, $\{6,7\})$  & $\lbrace2, 4\rbrace$ & $n-2$ & $n=2j+4$\\ 
\hhline{~----} 
& $N_{6.21}\equiv N_{4.4}\equiv N_{2.1} = (\emptyset, \emptyset, \{0,1,2,3,6,7\}$, $\{4,5\})$ & $\lbrace2, 4\rbrace$ & No & NA\\ 
\hhline{~----} 
& $N_{2.2} = (\{4,5\}$, $\{6,7\}, \emptyset, \{0,1,2,3\})$  & $\lbrace2, 4\rbrace$ & $n-2$ & $n=2j+4$\\ 
\hhline{~----} 
& $N_{6.53}\equiv N_{4.12}\equiv N_{2.3} = (\{0,1,2,3,6,7\}$, $\{4,5\}, \emptyset, \emptyset)$  & $\lbrace2, 4\rbrace$ & $n-2$ & $n=2j+4$\\ 
\hline 

\multirow{6}{*}{$3$} & $N_{5.21} \equiv N_{5.4}\equiv N_{3.1} = (\emptyset, \{0,1,2,3,6,7\}, \emptyset, \{4,5\})$  & $\lbrace3,5\rbrace$  & $n-1$ & $n=2j+4$\\ 
\hhline{~----} 
&  $N_{3.2} = (\emptyset, \emptyset, \{4,5,6,7\}$, $\{0,1,2,3\})$ & $\lbrace3, 5\rbrace$ & $n-1$ & $n=2j+4$\\ 
\hhline{~----} 
& $N_{4.7} \equiv N_{4.6} \equiv N_{3.3} = (\emptyset, \emptyset, \{0,1,2,3,4,5,6,7\}, \emptyset)$ & $\lbrace3,4\rbrace$ & No & NA\\
\hhline{~----}
& $N_{5.20} \equiv N_{5.5} \equiv N_{3.5} = (\emptyset, \{4,5\}, \emptyset,\{0,1,2,3,6,7\})$ & $\lbrace3,5\rbrace$ & $n-1$ & $n=2j+4$\\
\hhline{~----} 
& $N_{3.6} = (\{4,5,6,7\}$, $\{0,1,2,3\}, \emptyset, \emptyset)$ & $\lbrace3, 5\rbrace$ & No & NA \\ 
\hhline{~----} 
& $N_{4.15} \equiv N_{4.14} \equiv N_{3.7} = (\{0,1,2,3,4,5,6,7\}, \emptyset, \emptyset, \emptyset)$ & $\lbrace3,4\rbrace$ & No & NA\\ 
\hline 

\multirow{6}{*}{$4$} & $N_{4.2} = (\emptyset, \{4,5,6,7\}, \emptyset, \{0,1,2,3\})$ & $\lbrace4, 6\rbrace$ & $n-2$ & $n=2j+6$\\ 
\hhline{~----}
& $N_{5.27}\equiv N_{5.7}\equiv N_{5.6} \equiv N_{4.3} = (\emptyset, \{0,1,2,3,4,5,6,7\}, \emptyset, \emptyset)$ & $\lbrace4,5\rbrace$ & No & NA\\ 
\hhline{~----}
&  $N_{6.20} \equiv N_{4.5} = (\emptyset, \emptyset, \{4,5\}$, $\{0,1,2,3,6,7\})$  & $\lbrace4,6\rbrace$ & No & NA\\ 
\hhline{~----} 
&  $N_{4.10} = (\emptyset, \{0,1,2,3\}, \emptyset, \{4,5,6,7\})$ & $\lbrace 4, 6\rbrace$ & No & NA\\ 
\hhline{~----}
& $N_{5.23} \equiv N_{5.22}\equiv N_{5.11}\equiv N_{4.11} = (\emptyset, \emptyset, \emptyset, \{0,1,2,3,4,5,6,7\})$ & $\lbrace4,5\rbrace$ & No & NA\\
\hhline{~----} 
& $N_{6.52} \equiv N_{4.13} = (\{4,5\}$, $\{0,1,2,3,6,7\}, \emptyset, \emptyset)$ & $\{4,6\}$ & No & NA\\ 
\hline 

\multirow{2}{*}{$5$} & $N_{5.10} = (\emptyset, \emptyset, \{0,1,2,3\}$, $\{4,5,6,7\})$ & $\lbrace5, 7\rbrace$ & $n-1$ & $n=2j+6$\\ 
\hhline{~----} 
& $N_{5.26} = (\{0,1,2,3\}$, $\{4,5,6,7\}, \emptyset, \emptyset)$  & $\lbrace5, 7\rbrace$ & No & NA\\ 
\hline 
\end{tabular}
}
\end{table}
\end{example}

\subsection{Minimized Tree and Reversibility for all ${n \in \mathbb{N}}$}
\noindent Number of possible unique nodes for a CA is fixed and finite. So, the minimized reachability tree of a CA with any $n\ge n_0$, $n \in \mathbb{N}$, is always of finite height. (Here, while constructing minimized tree for a CA, our target is to draw the tree with all possible unique nodes in the CA, without considering the \emph{special} levels $n-\iota$, $1 \leq \iota \leq m-1$. This is because, these special levels can be identified from this minimized tree, as discussed below.) Therefore, a CA with any number of cells can also be represented by its minimized reachability tree of fixed height having only the unique nodes. This minimized tree is instrumental in relating reversibility of infinite CAs with that of finite CAs.

For a finite CA with fixed size $n$, the reversibility of the CA for that $n$ can be decided by using the minimized reachability tree. To decide this, we check whether any node at any level $l$, $1\leq l \leq n$, violates any reversibility conditions. To find the nodes of a level $p$, 
we take help of the loops in the tree.
For instance, if a node $\mathcal{N}$ has three levels associated with it-- $\lbrace i, i', i'' \rbrace$, $i''>i'>i$, then, it can be present in level $p$, if $(p-i) \equiv 0 \pmod {(i'-i)}$, or $(p-i) \equiv 0 \pmod {(i''-i)}$. Therefore, by this procedure, we can identify the nodes of level $n-\iota$, $1 \leq \iota \leq m-1$. A node can be present in level $n-\iota$, if $(n-\iota-i) \equiv 0 \pmod {(i'-i)}$, where $i' > i$ is any level associated with $\mathcal{N}$ and $1 \leq \iota \leq m-1$. However, if a node can exist in level $n-\iota$ ($1 \leq \iota \leq m-1$), Point~\ref{rtd5} of Definition~\ref{chap:semireversibility:def:tree} is applied to this node. So, to be reversible, these (modified) nodes also need to satisfy the reversibility conditions. Hence, we can verify the conditions of reversibility for any $n$.

However, as minimized reachability tree is same for any $n\ge n_0$, this tree contains the same information regarding reachability of the configurations for every $n \in \mathbb{N}$. So, for each $n\in \mathbb{N}$, one can check the minimized tree to find whether the CA is reversible or not. Nevertheless, this process of checking reversibility for every $n$ is not feasible. In this scenario, we exploit the property of the loops in the minimized reachability tree once again. Because, each loop invariably indicates the appearances of a specific node repeatedly at different levels in the tree, that is, for different $n$ in arithmetic series.

For a CA to be reversible (according to Definition~\ref{chap:semireversibility:def:rev}), every node of the minimized reachability tree has to satisfy the reversibility conditions of Corollary~\ref{chap:semireversibility:th:revcor1} and Corollary~\ref{chap:semireversibility:th:revcor2} for each $n \in \mathbb{N}$. In the minimized tree, if a node is not associated with any loop, this node can not appear in any other level except its present level. So, if all such nodes satisfy the reversibility conditions, we can proceed to other nodes associated with loop(s). Note that, to understand reversibility for any $n\in \mathbb{N}$, we need to observe these loops in the minimized tree. As we are to consider all $n\in \mathbb{N}$, each node associated with a loop can appear in each of the levels $n-\iota$, $1 \leq \iota \leq m-1$ for some $n$. So, every node associated with loop(s) is verified whether it violates any reversibility conditions of Corollary~\ref{chap:semireversibility:th:revcor1} and Corollary~\ref{chap:semireversibility:th:revcor2} for any of levels $n-\iota$, $1 \leq \iota \leq m-1$ by applying Point~\ref{rtd5} of Definition~\ref{chap:semireversibility:def:tree} on that node. If no such node exists in the minimized reachability tree, that means, for every $n \in \mathbb{N}$, the tree satisfies all reversibility conditions. That is, the CA is reversible for every $n \in \mathbb{N}$ (Definition~\ref{chap:semireversibility:def:rev}).
Therefore, by using minimized reachability tree for a CA, we can decide whether a CA is reversible (according to Definition~\ref{chap:semireversibility:def:rev}).

\begin{example}
Consider the $2$-state $4$-neighborhood CA $1010101010101010$. The minimized tree for this CA is shown in Figure~\ref{Chap:semireversible:fig:min_rt_rev}. The tree contains only $15$ unique nodes and the height of the minimized tree is $3$. Hence, for any $n\geq 3$, the minimized tree remains the same.
\begin{figure}[!h]
		 \centering
			\includegraphics[width= 4.5in, height = 2.3in]{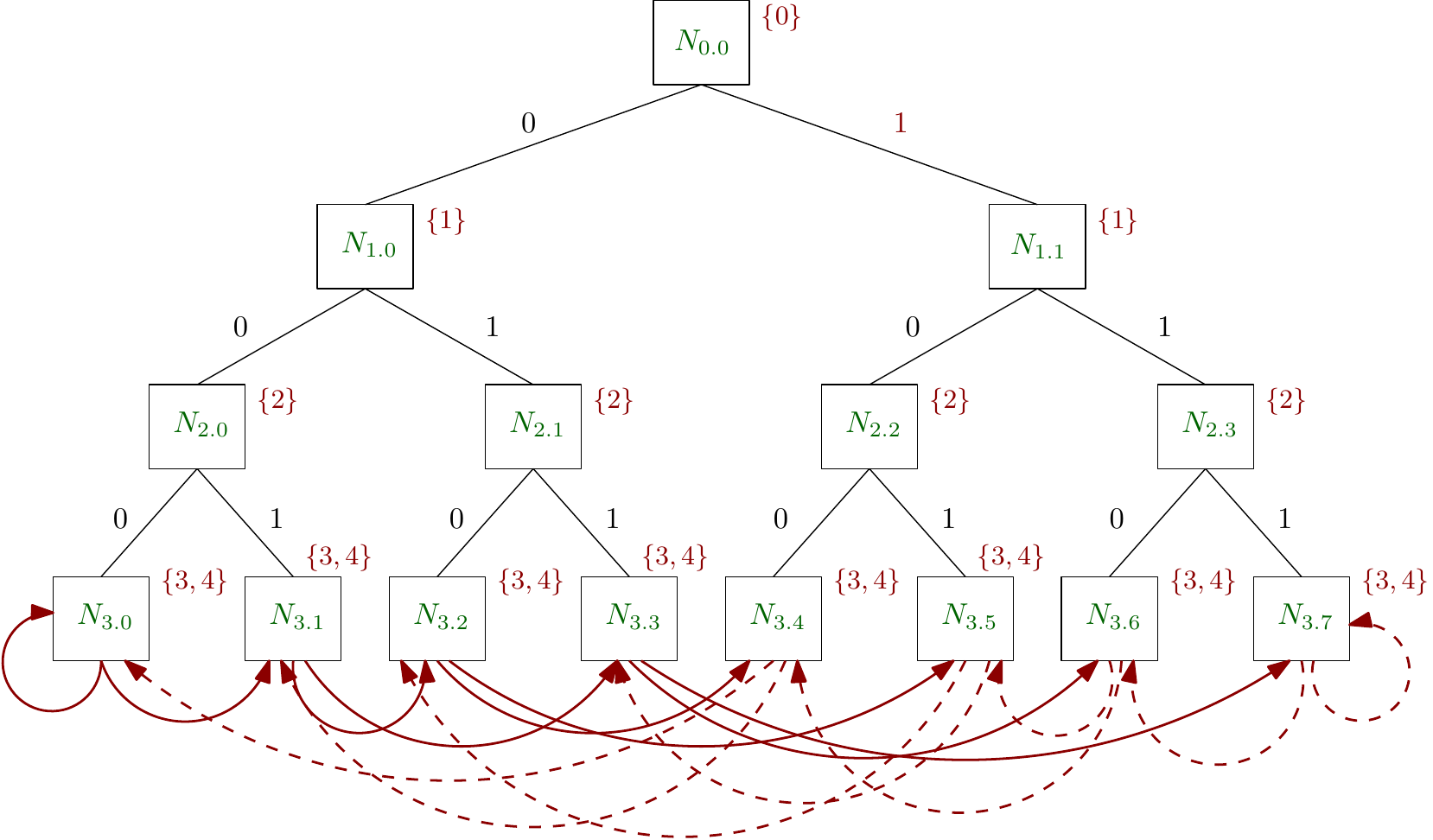}
			\caption{Minimized Reachability tree of $2$-state $4$-neighborhood CA with rule $1010101010101010$}
			\label{Chap:semireversible:fig:min_rt_rev}
	\end{figure}
For this CA, the nodes of the minimized tree satisfy every reversibility condition. Further, each of the nodes of level $3$ is associated with self-loop. But, even after applying Point~\ref{rtd5} of Definition~\ref{chap:semireversibility:def:tree} on those nodes, none of the modified nodes violate any reversibility conditions for any $n \in \mathbb{N}$. Hence, the CA is reversible for each $n \in \mathbb{N}$.
\end{example}

Moreover, the tree can help us to identify in which class of reversibility, the CA belongs to. The next section discusses about this.

\section{Minimized Reachability Tree and Semi-reversibility}\label{Chap:semireversibility:sec:semi_tree}
\noindent Using the minimized tree for a CA, we can understand the reversibility class of the CA. If a CA is strictly irreversible, it can be identified easily by using Corollary~\ref{Chap:semireversible:corollary:irreversibility}. For such a CA, two RMTs exist of form $(x,x\cdots,x)$, $x \in S$, for which the next state values are same. For instance, the $3$-neighborhood $3$-state CA $102012120012102120102102120$ has $R[13]=R[0]=0$, so the CA is strictly irreversible and there is no need to construct minimized reachability tree for the CA. Similarly, if a CA rule is unbalanced, but the CA is not strictly irreversible, it is trivially semi-reversible (by Corollary~\ref{chap:semireversibility:th:revcor3}). Take as example the $2$-neighborhood $3$-state CA $021122011$. The rule is unbalanced, so the CA is trivially semi-reversible (irreversible for all $n\ge 2$). Here also, we do not have to construct the minimized tree. However, for other cases, we need to construct the minimized reachability tree for that CA.

During the construction of the minimized  tree, we test the reversibility conditions of Corollary~\ref{chap:semireversibility:th:revcor1} and Corollary~\ref{chap:semireversibility:th:revcor2} at addition of each new node in the tree. If a reversibility condition is violated before completion of the construction, we conclude that the CA is trivially semi-reversible. For example, some nodes may be found, which are not part of any loop, but for which the CA does not satisfy the reversibility conditions. If a CA has such node(s) in the minimized tree at level $i$, that means, for every $n\geq i$ ($i\ge m$), the CA is irreversible. So, such CAs are trivially semi-reversible. (If we draw minimized reachability tree for a strictly irreversible CA, then also, reversibility conditions are violated before completion of the tree.) 
For example, for a $2$-neighborhood $3$-state CA $012122001$, the minimized tree has an unbalanced node $N_{1.0}=(\{3, 4, 5, 6, 7, 8\} ,\emptyset, \{6, 7, 8\})$ at level $1$. So, the CA is irreversible for all $n>1$, that is, it is a trivial semi-reversible CA.

Similarly, if a node has a self-loop at level $i$, that means, it will be present in the reachability tree for each $n\geq i$. If such a node violates conditions of reversibility after applying Point~\ref{rtd5} of Definition~\ref{chap:semireversibility:def:tree}, the CA is trivially semi-reversible. Whenever a CA is detected to be trivial semi-reversible, there is no need to continue construction of the minimized reachability tree. However, in case of non-trivial semi-reversible CA and reversible CA, we can complete construction of the minimized reachability tree. For the reversible CAs, no node of the minimized reachability tree violates the reversibility conditions. Nevertheless, we next discuss the way of identifying a non-trivial semi-reversible CA with the infinite set of sizes for which the CA is reversible.


A CA is non-trivially semi-reversible, if there exists infinite number of sizes for which it is reversible. In the minimized reachability tree, reversibility for any size $n$ can be confirmed if no node in the tree violates reversibility conditions for that $n$. However, if a CA is irreversible for some $n$, the minimized tree has node(s) which violate(s) reversibility conditions for those $n$. If a node at level $i$ violates any reversibility condition, that means the CA is irreversible for all $n>i$. Hence, it is trivially semi-reversible. Nonetheless, a node associated with loop(s) may violate these conditions after applying Point~\ref{rtd5} of Definition~\ref{chap:semireversibility:def:tree} to that node for some loop. This means, the CA is irreversible for all those $n$ in arithmetic series implied by the loop.

For example, say, according to the values of a loop, a node can exist in level $n-\iota$ ($1 \leq \iota \leq m-1$). So, by Point~\ref{rtd5} of Definition~\ref{chap:semireversibility:def:tree}, only $\frac{1}{d}$ of the RMTs of the original node of the loop remains valid for this node. If this (modified) node does not satisfy the reversibility conditions for levels $n-\iota$ ($1 \leq \iota \leq m-1$) (using Corollary~\ref{chap:semireversibility:th:revcor1} and Corollary~\ref{chap:semireversibility:th:revcor2}), the CA is irreversible for those values of $n$. Therefore, if a node does not satisfy these conditions, we can get an expression over cell length $n=k(i'-i)+i+\iota$, where $k \ge 0$ and $1 \leq \iota \leq m-1$, for which the CA is irreversible. This expression is termed as \textit{irreversibility expression}. For each loop violating the reversibility conditions for some values of $n$, we get an irreversibility expression. By using these irreversibility expressions, we can find the set of sizes for which the CA is irreversible. Let $\mathcal{I}$ be the set of CA sizes from the irreversibility expressions for which the CA is irreversible. Then, $\mathbb{N}\setminus\mathcal{I}$ denotes the set of lattice sizes for which the CA is reversible.


If the minimized reachability tree for a CA has some loops, where for some specific values of $n$, the nodes of the loops violate reversibility condition(s), then the CA is irreversible for only those specific values of $n$. For other $n$, the CA is reversible. Such a CA is non-trivially semi-reversible. Because, for such CAs, the set $\mathbb{N}\setminus\mathcal{I}$ denotes an infinite set of $n$ for which the CA is reversible. However, for trivial semi-reversible CAs, this set indicates only finitely many sizes for which the CA is reversible. Similarly, a CA is declared reversible, when
there is no $n$, for which an irreversibility expression can be found in the minimized tree. That is, the set $\mathcal{I} = \emptyset$.
\begin{example}
	Consider the minimized reachability tree of the ECA $75 (01001011)$ shown in Figure~\ref{fig:rt3}. Here, the CA violates reversibility conditions only when the lattice size is an even number (see Table~\ref{tab:ex:ECA75} for details). So, the irreversibility expressions are $n=2j+2$, $n=2j+4$ and $n=2j+6$, where $j \ge 0$. That is, the final expression of irreversibility is $n = 2j +2$, $j \ge 0$ and the set $\mathcal{I} = \{k ~|~ k\equiv 0\pmod{2}\}$. Therefore, ECA $75(01001011)$ is reversible for the set $\mathbb{N}\setminus\mathcal{I}=\{n ~|~ n \equiv 1 \pmod{2}\}$, that is, when $n$ is an odd number.
	
	Similarly, for the $2$-state $4$-neighborhood CA $1010101010101010$ of Figure~\ref{Chap:semireversible:fig:min_rt_rev}, the set $\mathcal{I} = \emptyset$. So, the CA is reversible for each $n\in \mathbb{N}$.
\end{example}

Therefore, the minimized reachability tree can segregate the non-trivial semi-reversible CAs from the trivial semi-reversible CAs and reversible CAs. In other words, given a CA, it can efficiently detect its reversibility class. Further, we can also deduce the expression of irreversibility for non-trivial semi-reversible CAs.

While conducting an experiment on all possible rules for some particular $d $ and $m$, we have observed that, for ECAs ($d=2$, $m=3$), maximum height of the minimized reachability tree according to 
 is $5$, for $2$-state $4$-neighborhood CAs, the height is $9$ and for $3$-state $3$-neighborhood CAs the maximum height of the minimized tree for deciding non-trivial semi-reversibility is $19 $. Table~\ref{Chap:semireversibility:tab:results} shows some sample results of this experiment. In this table, the first column represents the number of states per cell ($d$), second column represents the number of neighbors ($m$) and the third column shows a CA rule for that $d$ and $m$. The number of unique nodes generated in the minimized tree
for that CA is reported in the fourth column along with the height of the minimized reachability tree (Column $5$); whereas, the reversibility class of the CA with an expression of irreversibility (Column $7$) is depicted in the sixth column.


\begin{table}[!h]
\begin{center}
\caption{Classification of some sample rules using minimized reachability tree}
\label{Chap:semireversibility:tab:results}
\resizebox{1.0\textwidth}{!}{
\begin{tabular}{|c|c|c|c|c|c|c|}
\hline 
$d$ & $m$ & Rule & $M$ & Height of tree & Reversibility Class? & Expression of irreversibility\\ 
\hline 
$2$ & $ 3 $ & $ 01010101 $ & $ 7 $ & $ 2 $ & Reversible & $\emptyset$\\ 
\hline 
$ 2 $ & $ 3 $ & $ 00011110 $ & NA & NA & Strictly irreversible & $\forall n \in \mathbb{N}$\\ 
\hline 
$ 2 $ & $ 3 $ & $ 00101101$ & $ 21 $ &  $ 5 $ & Non-trivial semi-reversible & $n=2j+2$, $j \ge 0$\\ 
\hline 
$ 2 $ & $ 3 $ & $ 10010110 $ & $ 12 $ &  $ 4 $ & Non-trivial semi-reversible & $n=3j+3$, $j \ge 0$ \\ 
\hline 
$ 2 $ & $ 3 $ & $ 00101011 $ & $ 4 $ & $ 2 $ & Trivial semi-reversible & $n\ge 3$\\
\hline
$3$ & $ 3 $ & $ 012012012012012210012102012 $ & $ 84 $ & $ 6 $ & Non-trivial semi-reversible & $n=4j+4$, $j \ge 0$\\ 
\hline 
$ 3 $ & $ 3 $ & $ 012210210102012102210210012 $ & $1371$ & $19$ & Non-trivial semi-reversible &  $n=2j+4$, $n=3j+3$, $j \ge 0$\\ 
\hline 
$ 3 $ & $ 3 $ & $ 012012012012012012012012012 $ & $ 13 $ &  $ 2 $ & Reversible & $\emptyset$ \\ 
\hline 
$ 3 $ & $ 3 $ & $ 210201102201120210201012102 $ & NA & NA & Strictly irreversible & $\forall n \in \mathbb{N}$\\
\hline
$ 2 $ & $ 3 $ & $ 021101110202222202110010021 $ & $ 1345 $ & $ 19 $ & Non-trivial semi-reversible & $n=2j+4$, $n=3j+3$, $j \ge 0$\\
\hline
$3$ & $ 3 $ & $ 111011011222220122000102200 $ & $ 252 $ & $ 9 $ & Non-trivial semi-reversible &  $n=2j+4$, $j \ge 0$\\
\hline 
$3$ & $ 3 $ & $ 102120120102120021120120120 $ & $ 104 $ & $7 $ & Non-trivial semi-reversible &  $n=2j+4$, $j \ge 0$\\
\hline 
$ 2 $ & $ 4 $ & $ 0001000011101111 $ & $65$ & $7$ & Non-trivial semi-reversible & $n=3j+3$, $j \ge 0$\\ 
\hline 
$ 2 $ & $ 4 $ & $ 0101101010100101$ & $ 56 $ &  $ 9 $ & Non-trivial semi-reversible & $n=7j+7$, $j\ge 0$\\ 
\hline 
$ 2 $ & $ 4 $ & $ 0000111101001011 $ & $ 32 $ &  $ 5 $ & Reversible & $\emptyset$ \\ 
\hline 
$ 2 $ & $ 4 $ & $ 0000111101001110 $ & NA & NA & Strictly irreversible & $\forall n \in \mathbb{N}$\\
\hline
$ 2 $ & $ 4 $ & $ 0000111101010101 $ & $ 4 $ & $ 2 $ & Trivial semi-reversible & $n\ge 4$\\
\hline 
$ 2 $ & $ 4 $ & $ 0000111101001011 $ & $ 32 $ &  $ 5 $ & Reversible & $\emptyset$ \\ 
\hline 
$ 2 $ & $ 4 $ & $ 1101110010001101 $ & NA & NA & Strictly irreversible & $\forall n \in \mathbb{N}$\\
\hline
$ 2 $ & $ 4 $ & $ 1101110000110010 $ & $ 2 $ & $ 1 $ & Trivial semi-reversible & $n\ge 4$\\
\hline
\end{tabular} }
\end{center}
\end{table}

Hence, minimized reachability tree of size $5$, $9$ and $19$ are sufficient to decide reversibility class of ECAs, $4$-neighborhood $2$-state CAs and $3$-neighborhood $3$-state CAs respectively. That is, from a small set of sizes, reversibility class of the CAs can be identified. 

\section{New Relation about Reversibility}\label{Chap:semireversible:sec:remark}
\noindent Theorem~\ref{Chap:semi-reversibility:Th:strictirreversibility} and minimized reachability tree guide us to decide the nature of reversibility of any $1$-D CA. By construction, height of the minimized tree for a CA is finite, say $n_0$. So, we can deduce about the reversibility class of the CA, if we construct 
the reachability tree for each $n$ up to length $n_0+m-1$. Otherwise, we can construct 
the minimized reachability tree for an $n$-cell CA, where $n \ge n_0$ and detect the reversibility class of the CA for any $n \in \mathbb{N}$. Therefore, the sufficient length to decide the reversibility class of CA is $n_0+m-1$ for which the minimized reachability tree has height $n_0$ and contains all possible unique nodes for the CA. Now we can relate the different cases of reversibility using the global transition functions.

If a finite CA with rule $R$ is found reversible by minimized reachability tree (resp. strictly irreversible by Theorem~\ref{Chap:semi-reversibility:Th:strictirreversibility}), then the CA is also reversible (resp. strictly irreversible) for configurations of length $n$, for all $n \in \mathbb{N}$, under periodic boundary condition; that is, for the set of all periodic configurations. That means, for a CA with local rule $R$, injectivity of $G_n$ (Case $4$) for all $n \in \mathbb{N}$, implies injectivity of $G_P$ (Case $2$), and hence, injectivity of $G$ (Case $1$) and $G_F$ (Case $3$). We can further note the following:
\begin{itemize}

\item For a finite CA with a set of periodic configurations of length $n$ for a fixed $n$, $G_n$ injective $\Rightarrow$ $G_n$ surjective.

\item If a CA is bijective for periodic configurations of length $n$, for all $n$, it is also bijective for periodic configurations of length $n$, for a fixed $n$, but the converse is not true. That is, 
$G_P$ injective \stackunder{$\rightarrow$}{$\nleftarrow$}
$G_n$ injective.

\item If a CA is not bijective for periodic configurations of length $n$, for a fixed $n$, then it is not bijective for periodic configurations of length $n$, for all $n$, but the converse is not true. That is, $G_n$ not bijective \stackunder{$\rightarrow$}{$\nleftarrow$} 
$G_P$ not bijective.

\item If $G$ is injective, then $G_n$ is also injective for any $n\in \mathbb{N}$, but the converse is not true. However, if $G_n$ is not injective, then $G$ is also not injective.

\item The algorithms of Amoroso and Patt \cite{Amoroso72} as well as of Sutner \cite{suttner91}, report the trivial semi-reversible CAs as neither surjective nor injective, but the non-trivial semi-reversible CAs as not injective but surjective for infinite configurations.
\end{itemize}

\noindent 
By Definition~\ref{Chap:semireversible:def:nontriv_semi}, the finite CAs which are reversible for length $n$, for all $n$, are a special case of non-trivial semi-reversible CAs. However, for a finite CA with a local rule $R$, we can get different $G_n$ by varying the lattice size $n$. To relate the injectivity of $G_P$ with $G_n$, let us define the following:
\begin{definition}
Let, ${G_R}^* = \{G_n ~|~ n \in \mathbb{N}\}$ be the set of global transition functions for rule $R$, when $n \in \mathbb{N}$. ${G_R}^*$ is called \textbf{reversible} if $G_n$ is reversible for each $n \in \mathbb{N}$, \textbf{strictly irreversible} if $G_n$ is irreversible for each $n \in \mathbb{N}$ and \textbf{semi-reversible} if $G_n$ is reversible for some $n \in \mathbb{N}$.
\end{definition}

Let us now relate ${G_R}^*$ with $G$, $G_P$ and $G_F$.
We can understand that, when ${G_R}^*$ is reversible, that means, the finite CA is reversible for every $n \in \mathbb{N}$, hence $G_P$ is injective. Similarly, if ${G_R}^*$ is strictly irreversible, $G_P$ is not injective.

\begin{figure*}[hbt]
\centering
\includegraphics[width= 4.4in, height = 1.4in]{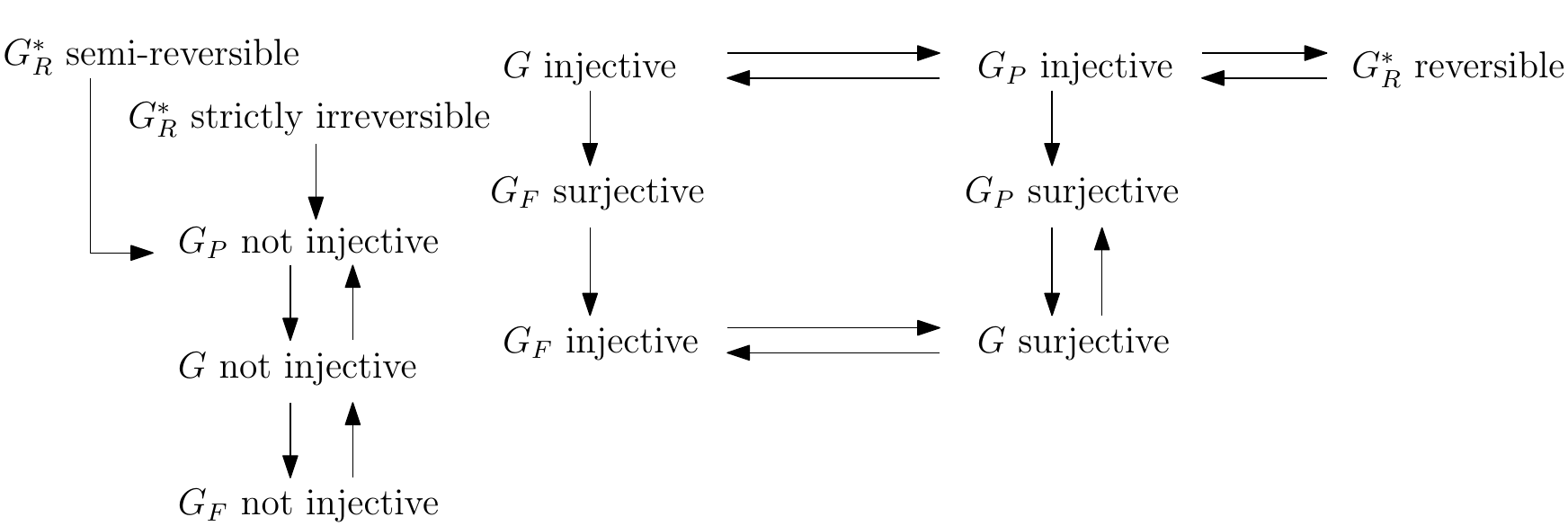}
\caption{Updated relationship among various kinds of reversibility of $1$-dimensional CAs}
\label{Chap:semireversible:fig:rev_rel_2}
\end{figure*}

 Let us define $S_R$ to be the set of CA rules which are reversible for some cell length $n\in \mathbb{N}$. So, all the semi-reversible and reversible CAs belong to this set $S_R$, that is $S_R = \{R ~|~ {G_R}^*$ is reversible $\lor ~{G_R}^*$ is semi-reversible\}. Let us further define $S_I$ to be the set of CA rules whose $G$ is bijective over the set of infinite configurations, $S_F$ to be the set of CA rules whose $G_F$ is bijective over the set of all finite configurations, and $S_P$ to be the set of CA rules whose $G_P$ is bijective over the set of periodic configurations. From the above discussions, it is evident that, $S_P$ is a subset of the set $S_R$.
 As, for one dimensional CAs, $S_F$, $S_I$ and $S_P$ are equivalent, so, $S_R$ is superset of all these sets. That is, $S_P\equiv S_I \equiv S_F \subset S_R$. 
Hence, Figure~\ref{Chap:semireversible:fig:rev_rel} is updated to depict the relation among the cases of reversibility in Figure~\ref{Chap:semireversible:fig:rev_rel_2}.

\noindent The above discourse points to the fact that, the cell length $n$ plays a salient role to determine the reversibility of the CA. Depending on the value of $n$, a CA can be reversible or irreversible. For each CA, we can get an $n_0$, height of the minimized tree, which is necessary to identify its reversibility class. However, is there any sufficient value of $n$, which is agreed by every CA, to predict the reversibility behavior of any CA? This question remains open for future research.\\

%

\section{Conclusion}
\label{Chap:semireversible:sec:con}
\noindent In this work, we have divided $1$-D CAs into three classes -- reversible, semi-reversible and strictly irreversible CAs. Semi-reversible CAs are further classified into two sub-classes -- trivially semi-reversible and non-trivially semi-reversible. Reachability tree has been used to identify the appropriate class of a CA and find the expression of irreversibility for any non-trivially semi-reversible CA. It is observed that, from the reversibility for a some small set of sizes, we can confer about the reversibility of the CA defined over periodic configurations, and consequently for infinite CAs. Finally, we have related the four cases of reversibility: infinite configurations with $G$, periodic configurations with $G_P$, finite configurations for all $n \in \mathbb{N}$ with $G_F$ and configurations for fixed $n$ with $G_n$ under periodic boundary condition.

However, we have observed that for a CA with rule $R$, the minimized reachability tree does not have much unique nodes. In fact, after certain number of levels, no new nodes are generated. Since, number of levels and number of cells are related, this observation raises the following question-- 
what is the tight upper bound of the necessary and sufficient lattice size $(n_0)$ to decide reversibility class of a finite CA? Efforts may be taken to answer this question. We have also noticed that, for the trivial semi-reversible CAs, we need not to complete the construction of minimized reachability tree. Hence, the question arises, is there any other criteria for identifying such CAs without constructing the minimized tree?
%
%

\section{Acknowledgments}
\noindent This research is partially supported by Innovation in Science Pursuit for Inspired Research (INSPIRE) under Dept. of Science and Technology, Govt. of India. 

\noindent The authors are grateful to Prof. Mihir K. Chakraborty, Visiting Professor, Department of Humanities and Social Sciences, Indian Institute of Engineering Science and Technology, Shibpur for his valuable suggestions.

\section*{Appendix}
\noindent Here, proofs of the reversibility theorems for reachability tree are depicted.
\begin{theorem}
The reachability tree of a finite reversible CA of length $n$ $(n \geq m)$ is complete.
\end{theorem} 

\begin{proof} If a CA is reversible for length $n$, every configuration of it is reachable. So, in the corresponding reachability tree of $n+1$ levels, the number of leaves is $d^n$. Hence, the tree is complete.
\end{proof}

\begin{theorem}
The reachability tree of a finite CA of length $n$ $ (n \geq m)$ is complete if and only if
\begin{enumerate}[topsep=0pt,itemsep=0ex,partopsep=2ex,parsep=1ex]
\item \label{c2}  The label $l_{n-\iota.j}$, for any $j$, contains only $d^{\iota-1}$ RMTs, where $1 \leq \iota \leq m-1$; that is, \[\mid \bigcup\limits_{0 \leq k \leq d^{m-1} -1} {\Gamma_{k}^{E_{n-\iota.j}}}\mid = d^{\iota-1}\]

\item \label{c3} Each other label $l_{i.j}$ contains $d^{m-1}$ RMTs, where $ 0 \leq i \leq n-m$; that is, \[ \mid \bigcup\limits_{0 \leq k \leq d^{m-1} -1} {\Gamma_{k}}^{E_{i.j}}\mid = d^{m-1}\]
\end{enumerate}
\end{theorem}

\begin{proof}
\begin{description}[leftmargin=0pt]
\item\noindent\underline{\textit{If Part:}} Let us consider, the number of RMTs in the label of an edge is less than that is mentioned in (\ref{c2}) and (\ref{c3}). This implies,

(i)\label{i1} The label $l_{n-1.j}$ is empty, for some $j$. That is, $ \bigcup\limits_{0 \leq k \leq d^2 -1} {\Gamma_{k}^{E_{n-1.j}}}= \emptyset $. So, the tree has a non-reachable edge and it is incomplete.

(ii) \label{i2} In general, for some $j$, the label $l_{n-\iota.j}$ contains less than $d^{\iota-1}$ RMTs, where $1\leq \iota \leq m-1$. That is, $ \mid \bigcup\limits_{0 \leq k \leq d^{m-1} -1} {\Gamma_{k}^{E_{n-\iota.j}}}\mid \leq d^{\iota-1}-1$. Then, the number of RMTs in the node $N_{n-\iota+1.j} \leq d(d^{\iota-1}-1)$. However, according to Point~\ref{rtd5} of the Definition~\ref{chap:semireversibility:def:tree}, only $\frac{1}{d}$ of the RMTs of these levels are valid. So, the number of valid RMTs is $ \leq \frac{d(d^{\iota-1}-1)}{d} = (d^{\iota-1}-1)$. If this scenario is continued for all $\iota$, then in the best case, the tree may not have any non-reachable node up to level $(n - 1)$. But, at level $n-1$, there will be at least one node for which $ \mid \bigcup\limits_{0 \leq k \leq d^{m-1} -1} {\Gamma_{k}^{N_{n-1.g}}}\mid \leq d-1$, that is, the maximum number of possible edges from this node is $d-1$. Hence, at least one (non-reachable) edge ${E_{n-1.b}}$ exist in the tree for which $\bigcup\limits_{0 \leq k \leq d^{m-1} -1} {\Gamma_{k}^{E_{n-1.b}}}= \emptyset $.

(iii) \label{i3} Let, every other label $l_{i.j}$ contains less than $d^{m-1}$ RMTs, that is, \\$ \mid \bigcup\limits_{0 \leq k \leq d^{m-1} -1} {\Gamma_{k}^{E_{i.j}}}\mid < d^{m-1}$, $(0 \leq i \leq n-m)$. Then, $ \mid \bigcup\limits_{0 \leq k \leq d^{m-1} -1} {\Gamma_{k}^{N_{i+1.j}}}\mid < d^m$. The corresponding node $N_{i+1.j}$ may have $d$ number of edges. Here also, in best case, the tree may not have any non-reachable edge up to level $(n - 1)$. Then there exists at least one
 node $N_{n-1.p}$ where $ \mid \bigcup\limits_{0 \leq k \leq d^{m-1} -1} {\Gamma_{k}^{N_{n-1.p}}}\mid < d^2$. Since the node is at level $(n - 1)$, it has maximum $\frac{(d^2-1)}{d} < d$ valid RMTs (see Point~\ref{rtd5} of the Definition~\ref{chap:semireversibility:def:tree}). Therefore, there exists at least one edge from the node $N_{n-1.p}$, where $ \mid \bigcup\limits_{0 \leq k \leq d^{m-1} -1} {\Gamma_{k}^{E_{n-1.q}}}\mid =\emptyset$, which makes the tree an incomplete one by (i).

On the other hand, any intermediate edge ${E_{i.j_1}}$ has, say, more than $d^{m-1}$ RMTs, that is, $ \mid \bigcup\limits_{0 \leq k \leq d^{m-1} -1} {\Gamma_{k}^{E_{i.j_1}}}\mid$ $\geq d^{m-1}$. Then there exists an edge $ E_{i.j_2}$ in the same label $i$ for which \\$ \mid \bigcup\limits_{0 \leq k \leq d^{m-1} -1} {\Gamma_{k}^{E_{i.j_2}}}\mid < d^{m-1}$, where $ 0 \leq i \leq n-m$, and $j_1 \neq j_2$. So, by (iii), the tree is incomplete. Now, if for any $p$, label $l_{n-\iota.p}$ contains more than $d^{\iota-1}$ RMTs, then also there exists an edge $E_{n-\iota.q}$ for which $ \mid \bigcup\limits_{0 \leq k \leq d^{m-1} -1} {\Gamma_{k}^{E_{n-\iota.q}}}\mid$ $< d^{\iota-1}$. Hence, the tree is incomplete (by (ii)).
Hence, if the number of RMTs for any label is not same as mentioned in (\ref{c2}) and (\ref{c3}), the reachability tree is incomplete. 

\item\noindent\underline{\textit{Only if Part:}} Now, consider that, the reachability tree is complete. The root $N_{0.0}$ has $d^m$ number of RMTs. In the next level, these RMTs are to be distributed such that the tree remains complete. Say, an edge $E_{0.j_1}$ has less than $d^{m-1}$ RMTs, another edge $E_{0.j_2}$ has greater than $d^{m-1}$ RMTs and other edges $E_{0.j} (0 \leq j,j_1,j_2 \leq d-1$ and $j_1 \neq j_2 \neq j)$ has $d^{m-1}$ RMTs. Then node $N_{1.j_1}$ has less than $d^m$ RMTs, $N_{1.j_2}$ has greater than $d^m$ RMTs and other edges $N_{1.j}$ has $d^m$ RMTs. There is no non-reachable edge at level $1$. 
Now, this situation may continue up to level $n-m$. However, at level $(n-\iota)$, $1\leq \iota \leq m-1$, only $\frac{1}{d}$ of the RMTs are valid (see Definition~\ref{chap:semireversibility:def:tree}). So, the nodes at level $n-m+1$ with less than $d^m$ RMTs has at maximum $d^{m-1}-1$ valid RMTs, nodes at level $n-m+2$ has $<d^{m-2}$ valid RMTs and so on. So, at level $n-1$, there exists at least one node $N_{n-1.p}$, such that $ \mid \bigcup\limits_{0 \leq k \leq d^{m-1} -1} {\Gamma_{k}^{N_{n-1.p}}}\mid < d$ for which the tree will have non-reachable edge (by (ii)). Similar situation will arise, if any number of intermediate edges have less than $d^{m-1}$ RMTs. Because, this means, some other edges at the same level have more than $d^{m-1}$ RMTs. Hence, the tree will be incomplete which contradicts our initial assumption. So, for all intermediate edges $E_{i.j}$, $ \mid \bigcup\limits_{0 \leq k \leq d^{m-1} -1} {\Gamma_{k}^{E_{i.j}}}\mid = d^{m-1}$, where $ 0 \leq i \leq n-m$.

Now, if this is true, then in general, at level $n-\iota$, the nodes have $d^{\iota}$ valid RMTs and every edge has $d^{\iota-1}$ valid RMTs. If an edge $E_{n-\iota.p}$ has less than $d^{\iota-1}$ RMTs, then the node $N_{n-\iota.p}$ has at maximum $d(d^{\iota-1}-1)$ RMTs out of which only $d^{\iota-1}-1$ are valid. Hence, in this way, at level $n-1$, we can get at least one edge from $N_{n-1.q}$, which is non-reachable making the tree incomplete. So, for the tree to be complete, each edge label $l_{n-\iota.j}$, for any $j$, is to have $d^{\iota-1}$ RMTs. Hence the proof.
\end{description}\end{proof}

\begin{corollary}
The nodes of a reachability tree of a reversible CA of length $n$ $(n \geq m)$ contains
\begin{enumerate}[topsep=0pt,itemsep=0ex,partopsep=2ex,parsep=1ex]
\item $d$ RMTs, if the node is in level $n$, i.e. $ \mid \bigcup\limits_{0 \leq k \leq d^{m-1} -1}{\Gamma_{k}^{N_{n.j}}} \mid = d$ for any $j$.

\item $d^\iota $ RMTs, if the node is at level $n-\iota$ i.e, $ \mid \bigcup\limits_{0 \leq k \leq d^{m-1} -1}{\Gamma_{k}^{N_{n-\iota.j}}} \mid = d^\iota$ for any $j$, where $1 \leq \iota \leq m-1$.

\item $d^m$ RMTs for all other nodes $N_{i.j}$, $ \mid \bigcup\limits_{0 \leq k \leq d^{m-1} -1}{\Gamma_{k}^{N_{i.j}}} \mid = d^m$ where ${ 0 \leq i \leq n-m}$.
\end{enumerate}
\end{corollary}

\begin{proof}This can be directly followed from Theorem~\ref{chap:semireversibility:th:revth2}.
\end{proof}

\begin{corollary} 
The nodes of the reachability tree of an $n$-cell $(n \geq m)$ reversible CA are balanced.
\end{corollary}

\begin{proof}For an $n$-cell $(n \geq m)$ reversible CA, the reachability tree is complete. So, each node has $d$ number of edges. Each node $N_{i.j}$ contains $d^m$ RMTs when $i < n-\iota$, $1 \leq \iota \leq m-1$ (Corollary ~\ref{chap:semireversibility:th:revcor1}) and each edge $E_{i+1.k}$, for any $k$, contains $d^{m-1}$ RMTs (Theorem~\ref{chap:semireversibility:th:revth2}). So, the node $N_{i.j}$ is balanced and the number of RMTs per each state is $d^{m-1}$. Similarly, the nodes of level $n-\iota$ are balanced. 
\end{proof}

\begin{corollary}
Let $R$ be the rule of a CA. Now, the CA is trivially semi-reversible (irreversible for each $n\geq m$) if the following conditions are satisfied--
\begin{enumerate}[topsep=0pt,itemsep=0ex,partopsep=2ex,parsep=1ex]
\item $R$ is unbalanced and
\item no two RMTs $r$ and $s$ exist, such that $r=(x,x\cdots,x)$ and $s=(y,y,\cdots,y)$, $x,y \in S$ and $R[r]=R[s]$.
\end{enumerate}
\end{corollary}

\begin{proof}If the rule $R$ is unbalanced, number of RMTs per each state is unequal. So, for every $n\ge m$, the root node $N_{0.0}$ of the corresponding reachability tree is unbalanced. Hence, there exists at least one edge $E_{0.j}$ where $\mid\bigcup\limits_{0\leq k \leq d^{m-1}-1} \Gamma_k^{E_{0.j}}\mid < d^{m-1}$ $(0 \leq j \leq d^{m-1}-1)$. Therefore, by Theorem~\ref{chap:semireversibility:th:revth2}, the CA is irreversible for every $n\ge m$.

However, the second condition ensures that, the CA does not violate Corollary~\ref{Chap:semireversible:corollary:irreversibility}. So, it is not irreversible for $n=1$. 

Therefore, the CA is reversible for a finite number of lattice sizes (for at least $n=1$) and irreversible for an infinite set of sizes (for each $n \ge m$). Hence, by Definition~\ref{Chap:semireversible:def:triv_semi}, the CA is trivially semi-reversible.
\end{proof}


 \bibliographystyle{elsarticle-num} 
\bibliography{References_thesis}
\end{document}